\documentclass[twocolumn,dvipsnames,nofootinbib,floatfix,groupedaddress,superscriptaddress]{revtex4-2}

\newcommand{\dec}{\text{dec}}

\usepackage[utf8]{inputenc}
\usepackage{xcolor,soul}
\usepackage{siunitx}
\usepackage{amsmath}
\usepackage{amssymb,amsthm,amsfonts} 
\usepackage{booktabs}
\usepackage{bbold}
\usepackage{paralist}
\usepackage{xcolor}
\usepackage{graphicx}
\usepackage[T1]{fontenc}
\usepackage{braket} 
\usepackage[normalem]{ulem}

\usepackage[colorlinks,bookmarks=false,linkcolor=blue,urlcolor=blue,citecolor=blue]{hyperref}

\usepackage{feynmp}
\usepackage{tikz-feynman}
\tikzfeynmanset{compat=1.1.0}

\begin{document}
\title{High-energy neutrino signals from supernova explosions:\\ a new window into dark photon parameter space}
\author{Vsevolod Syvolap}
\affiliation{Instituut-Lorentz for Theoretical Physics,
Leiden University, 2333 CA Leiden, Netherlands}
\author{Oleg Ruchayskiy}
\affiliation{Niels Bohr Institute, Copenhagen University,
Jagtvej 155A, Copenhagen, DK-2100, Denmark}
\begin{abstract}
Dark photons, hypothetical feebly interacting massive vector bosons, appear in many extensions of the Standard Model. This study investigates their production and subsequent decay during supernova explosions.
We demonstrate that the decay of dark photons, with masses ranging from 200 to 400 MeV, can lead to the emission of neutrinos with energies surpassing those emitted by supernovae.
These neutrinos therefore serve as a distinct signal of new physics, allowing for the exploration of previously uncharted regions of the dark photon parameter space and complementing both accelerator-based searches and other astrophysical constraints.
The signal is largely unaffected by the specifics of the supernova's temperature and density radial profiles outside the SN core, rendering the prediction both robust and model-independent.
Our results indicate that searching for high-energy neutrinos accompanying supernova explosions provides a novel approach to probe physics beyond the Standard Model, including dark photons, heavy neutral leptons, and other feebly interacting particles with masses in the hundreds of MeV range.
\end{abstract}
\maketitle

\section{Introduction}
\label{sec:intro}
The cores of exploding supernovae (SNe) can reach high matter densities ($\rho \sim 10^{14} $ g/cm$^3$) and temperatures ($T \sim \SI{30}{MeV}$), see e.g.~\cite{Janka:2017vlw,Alsabti:2017ahu}.
These astrophysical environments serve therefore as unique sites of copious productions of hypothetical particles that interact with matter ``weaker than neutrinos''.
Such particles are collectively known as \emph{feebly interacting particles} or \emph{FIPs}, see  \cite{Alekhin:2015byh,Lanfranchi:2020crw,Agrawal:2021dbo,Antel:2023hkf} for the recent overviews of their properties and of models that predict them. 
A commonly used approach to constrain feebly interacting particles with supernovae utilizes the fact, that SN right after the explosion is cooled primarily via the active neutrino emission which lasts for approximately 10 seconds. In the presence of additional cooling channels, like the FIPs emission, this duration might be shortened significantly \cite{Raffelt:1996wa}  which would contradict the SN1987a observations \cite{Hirata:1987hu,1987ESOC...26..229S,1987ESOC...26..237A}. Using the energy-loss argument different FIPs have been studied in the past \cite{DelaTorreLuque:2023huu,Hoof:2022xbe,Dev:2020eam,Chen:2022kal,Magill:2018jla,Heurtier:2016otg,Kazanas:2014mca,Chang:2016ntp,Hook:2021ous,Dent:2012mx}.
However,  improving existing bounds necessitates detailed SN explosion models that incorporate how neutrino emissions are influenced by any additional cooling channels - \cite{Burrows:1988ah} and is highly model-dependent, given the current level of understanding of details of SN explosions, c.f.\ \cite{Burrows:2020qrp}.

Given the advancement of neutrino experiments, one expects a plethora of observational data, should a nearby supernova explode.
This warrants new studies about more detailed constraints on the properties of FIPs, based on the detailed measurements of SN neutrino signal. This work has already started \cite{Fiorillo:2022cdq,Akita:2022etk,Syvolap:2023trc}.

This work extends the supernova constraints on the properties of \emph{dark photons} (DP) --- massive vector bosons that interact with the Standard Model sector exclusively through {kinetic mixing} with photons. The Lagrangian of the model is given by
\begin{equation}
    \mathcal{L}_{\text{DP}} = \mathcal{L}_{\text{SM}} -\frac14 F'_{\mu\nu} {F'}^{\mu\nu} + \frac {m_{A'}^2}2 A'_\mu {A'}^\mu -  \frac{e}2 U F^{\mu \nu} F'_{\mu \nu}
\end{equation} 
where $\mathcal{L}_{\text{SM}}$ is the Lagrangian of the Standard Model, $e$ is the electric charge, $U$ is the dimensionless kinetic mixing parameter,\footnote{Often called $\varepsilon$ in the literature} $F^{\mu \nu}$ and $F'^{\mu \nu}$ are field strength tensors, associated with the SM photon field and the dark photon respectively (original references can be found in \cite{Okun:1982xi,Holdom:1985ag}, and the current model status and experimental searches are reviewed in \cite{Fabbrichesi:2020wbt,Caputo:2021eaa}). Transitioning to the mass basis means that electrically charged Standard Model particles inherit an interaction with the dark photon, carrying a ``charge'' of $U\times e$. This mechanism enables the potential production and detection of dark photons.
Previously the production of dark photons in SN interiors and their impact on the SN cooling have been studied in ~\cite{Kazanas:2014mca,Chang:2016ntp,Hook:2021ous,Dent:2012mx}.  These works considered the additional SN cooling channel via dark photons and derived the limits based on the energy-loss arguments. 
An alternative test for the presence of dark photons can be the search for a detectable gamma-ray signal that dark photons would produce when decaying to the Standard model photons \cite{DeRocco:2019njg}. 

The present study focuses on the potential for detecting signals induced by dark photons using neutrino detectors, should a nearby Galactic supernova explode.
We propose a method similar to our previous work~\cite{Syvolap:2023trc}, see also \cite{Mastrototaro:2019vug,Akita:2022etk,Fiorillo:2022cdq} -- the detection of high-energy neutrinos accompanying the SN neutrino signal.
In contrast to the case of heavy neutral leptons explored in \cite{Mastrototaro:2019vug,Syvolap:2023trc}, dark photons do not have a sizeable direct decay into neutrinos.
However, dark photons with masses $m_{A} > 2m_\mu$ ($2m_\pi$), produced within the SN core, can decay into pairs of muons (correspondingly, charged pions), which in turn will decay into neutrinos.
These decays will occur in less dense regions of the supernova, allowing the resulting neutrinos to escape and be detected by current and future neutrino detectors, including the Hyper-Kamiokande (\emph{Hyper-K}) neutrino telescope \cite{Hyper-Kamiokande:2016srs}, JUNO \cite{JUNO:2021vlw}, and DUNE \cite{DUNE:2021tad}.
Given that dark photons are primarily produced in the core, where temperatures are significantly higher than that of the neutrinosphere \cite{Janka:2017vlw}, the energy of neutrinos resulting from dark photon decays will be noticeably higher than that of the typical neutrino signal, making these signals distinguishable.

 \begin{compactitem}[--]
    \item[] The paper is organized as follows.
    \item  In the section \ref{sec:DP_in_SN},  the production of the dark photons are described and their spectra are calculated. 
    \item In Section~\ref{sec:Dp_decays} we compute the distribution function of the DP's decay products -- muons and pions, the sources of the secondary neutrinos. Here we also analyze their thermalization happening before decay, which modifies the particles' spectra.
    \item In Section~\ref{sec:DP_secondary_neutrinos}, the decays of remaining muons and pions leading to the neutrinos as secondary particles are discussed and the neutrino spectra are computed. 
    \item The propagation and flavor content of the secondary neutrino flux is described in Section \ref{sec:oscillation}.
    \item Section~\ref{sec:detection} discusses detection by the secondary neutrinos by Hyper-K and other neutrino detectors. Our main result is then summarized in Figure~\ref{fig:Main_Result}. 
    \item In the final Section  \ref{sec:conclusion}, we discuss the results and future prospects. 
 \end{compactitem}

\section{Dark photons in the supernova core}
\label{sec:DP_in_SN}
In the supernova medium the production of the dark photon occurs via scatterings of charged particles.
The main production channel is the proton-nucleon bremsstrahlung:
\begin{equation}
  \label{eq:hadronic_production}
  p + N \to p + N + A'\;.
\end{equation}
(here $N$ is neutron or another proton).
DP can also be produced in the scatterings, involving charged leptons:
\begin{equation}
  \label{eq:leptonic_production}
    e^{\pm} + X \to e^{\pm} + X + A'
\end{equation}
where $X$ is either a proton or another charged lepton. 
Since the nucleon number density is higher than that of the leptons and the cross-section of strong interactions is higher, than the electromagnetic ones, the channels~(\ref{eq:leptonic_production}) are subdominant and will be neglected in our calculations.

The period after the core bounce, when the temperature and density in the core are the highest, leads to the most efficient DP production.
A proper analytical description of such a process is missing.
To simplify the calculation (and to make our results readily reproducible) we will employ a simplified model of the SN core -- a  sphere with a uniform baryon density $\rho_B = \SI{3e14}{g/cm^3}$ and the temperature $T  = \SI{30}{MeV}$.
We take both quantities as constant for the time duration  $t_{\text{emm}} = 10$ sec.
The most intense production of the dark photons photons occurs inside the core and our result  is not sensitive to the temperature/density profile outside (see also discussion below, in Section~\ref{sec:DP_secondary_neutrinos}).

To find the dark photon spectra $f_{A'}$, we use the Boltzmann equation:
\begin{widetext}
\begin{equation}
\label{eq:Boltzmann}
    \frac{df_{A'} (p_{A'})}{dt} = \int \prod_{i=1,2,3,4} \frac{d^3p_i}{(2\pi)^32E_i} S|\mathcal{M}|^2 (2\pi)^5\delta(p_3 + p_4 + p_{A'} - p_1 - p_2) f_1f_2(1-f_3)(1-f_4)(1+ f_{A'}).
\end{equation}
\end{widetext}
where $p_i$ are the momenta of initial and final nucleons; $S|\mathcal{M}|^2$
-- matrix element of the processes~\eqref{eq:hadronic_production} with the
symmetry factor $S$ included in it; and $f_{i}$'s are the phase-space
distribution functions of the nucleons.
We solve Eq.~\eqref{eq:Boltzmann} following the approach of \cite{Dent:2012mx,Kazanas:2014mca}. 
For scattering energies, that are small compared to the nucleon mass, the
integration over the 12-dimensional space can be considerably reduced, leading
to the 3-dimensional integral:
\begin{widetext}
\begin{equation}
  \frac{dN_{A'}}{dE} = 
  e^2 U^2 \alpha_\pi^2 n_i n_j \sqrt{\frac{T^3}{8\pi m_{N}^3}}\int du dv dz \sqrt{u v} e^{-u} \sqrt{1-\frac{q^2}{x^2}}\delta (u-v-E/T)\mathcal{I}_{ij}.
\end{equation}
\end{widetext}
where  
$\alpha_\pi \approx (2m_N/m_\pi)^2/4\pi$ is the pion-nucleon coupling \cite{Ericson:2000md};  $n_i,n_j$ are the number densities of 
nucleons, $u,v,z,q$ --
dimensionless variables, related to momenta and angles in the scattering
process.
Finally the quantities $\mathcal{I}_{ij}$ are the expression, different for pp and pn processes, and discussed further in Appendix \ref{app:Integrands}.

\begin{figure}
    \centering
    \includegraphics[width=0.5 \textwidth]{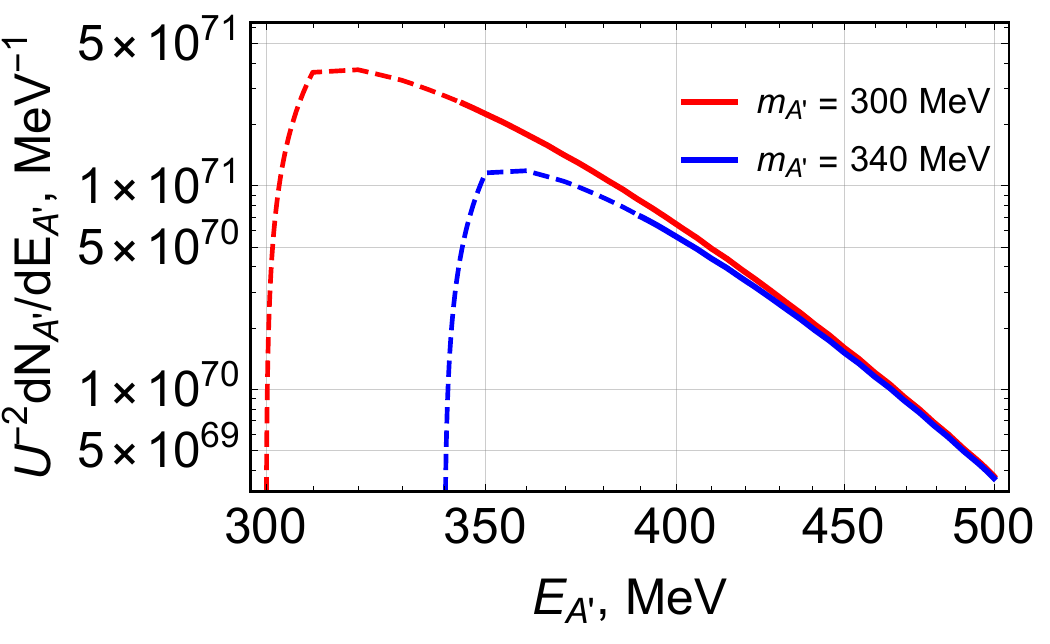}
    \caption{The energy spectra of dark photons, generated in the supernova core during the initial 10~sec post-bounce. Parts of the spectra represented by dashed lines are excluded due to gravitational trapping (see the text for details).}
    \label{fig:dark_photon_spectra}
\end{figure}

A low energy fraction of these dark protons will be gravitationally trapped inside the SN core, where they will decay and all their decay products will eventually thermalize, not affecting the SN neutrino spectra. 
The relevant part of the energy spectrum of dark photons is therefore given by: 
\begin{equation}
    \frac{dN_{A'}}{dE_{A'}} \to \frac{dN_{A'}}{dE_{A'}} \cdot \theta (E_{A'} - E_{g} - m_{A'}) 
\end{equation}
where the gravitational binding energy is defined as
\begin{equation}
    \label{eq:E_g}
    E_{g} = \frac{r_g}{R_c} m_{A'}
\end{equation}
with $R_c \simeq 10$~km is the typical radius of the SN core, and $r_g$ -- its Schwartzhield radius. 
The examples of the dark photon spectra $dN_{A'}/dE_{A'}$ produced during $t_{\text{emm}} \sim \SI{10}{sec}$ are presented in Fig.~\ref{fig:dark_photon_spectra}.

\section{Dark photon decays}
\label{sec:Dp_decays}
\begin{figure}
    \centering
    \includegraphics[width=0.5\textwidth]{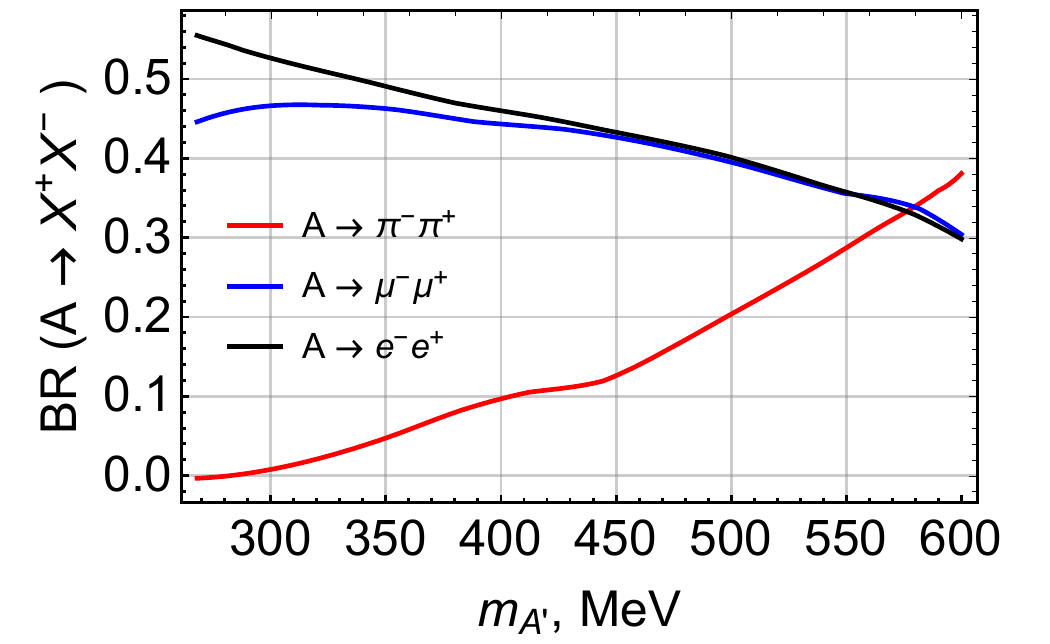}
    \caption{The branching ratios for the dominant dark photon decay modes. For masses of interest, no other decay channels with the branching fractions $\ge 1\%$ level exist.}
    \label{fig:DP_branchings}
\end{figure}

Our study concentrates on the parts of the dark photon parameter space where these particles decay \emph{inside} the supernova (but of course, outside of the supernova core).
Along with their decay, dark photons can get re-absorbed in the supernova environments. 
It has been shown in \cite{Dent:2012mx} that for dark photons with $m_{A'} > 2 m_\mu $ their decays are dominant for the depletion of the dark photon flux.
For masses $m_{A'} \lesssim \SI{600}{MeV}$ there are three main decay channels \cite{Fabbrichesi:2020wbt}:
\begin{eqnarray}
  A' &\rightarrow& e^- + e^+ \nonumber\\
  A' &\rightarrow& \mu^- + \mu^+ \label{eq:DP_decays}\\
  A' &\rightarrow& \pi^- + \pi^+ \nonumber 
\end{eqnarray}
The decay width into the lepton pair is given by
\begin{equation}
  \label{eq:Gamma_lept}
    \Gamma_{A'\rightarrow l^+ l^-} = \frac{\alpha}{3} U^2 m_{A'}\sqrt{\left(1-\frac{4m_l^2}{m_{A'}^2}\right)}\left(1+\frac{2m_l^2}{m_{A'}^2}\right)
\end{equation}
where $m_l$ is the mass of the relevant lepton. The decay width into hadrons is related to it as
\begin{equation}
  \Gamma_{A'\rightarrow \text{hadrons}} = \Gamma_{A'\rightarrow \mu^+ \mu-}
  \left(\frac{\sigma_{e^+e^-\rightarrow\text{hadrons}}}
    {\sigma_{e^+e^- \to \mu^+\mu^-}}\right)\biggr|_{\sqrt{s} = m_{A'}}
\end{equation}
where $\sigma_{e^+e^-\rightarrow\text{hadrons}}$ is the inclusive inelastic cross-section.
The branching ratios for these channels are shown in Fig.~\ref{fig:DP_branchings}.
The electron decay channel is not interesting to us as they are stable, and can not lead to the production of neutrinos,
 but muons and pions are long-lived and have neutrinos as their decay products.
The decay of these particles can therefore lead to delayed neutrino emission -- the subject of this study.

The energy spectra of the decay products have been calculated before \cite{Oberauer:1993yr} and for the 2-particle decay scenario has the form:
\begin{equation}
  f_{x}(E) = \frac{m_{A'}}{2\bar{E}}\text{Br}_{A \to \mu/\pi} \int\limits_{\text{E}_{\text{min}}(E)}\frac{d E_{A'}}{p_{A'}}\frac{dN_{A'}}{dE_{A'}}
  \label{eq:2-body_master_mu}
\end{equation}
The subscript $x$ stands for $\mu/\pi$, $\bar{E} \equiv m_{A'}/2$ is the energy of the decay product in the rest frame of the parent particle,
$p_{A'}$ is the absolute value of the DP's 3-momentum; and the quantity $\text{E}_{\text{min}}$ shows the minimum  values of the DP energy, that can produce a pion/muon with the energy $E$ in the laboratory frame: 
\begin{equation}
  \text{E}_{\text{min}}(E) = \max\left(E_g, E + \frac{m_{A'}^2}{4 E}\right)    
    \label{eq:Emin_eqn}
\end{equation}
In both decay processes the kinematics are the same, therefore the calculation will be the same up to the change in the product's mass.

\begin{figure}
    \centering
    \includegraphics[width=0.5\textwidth]{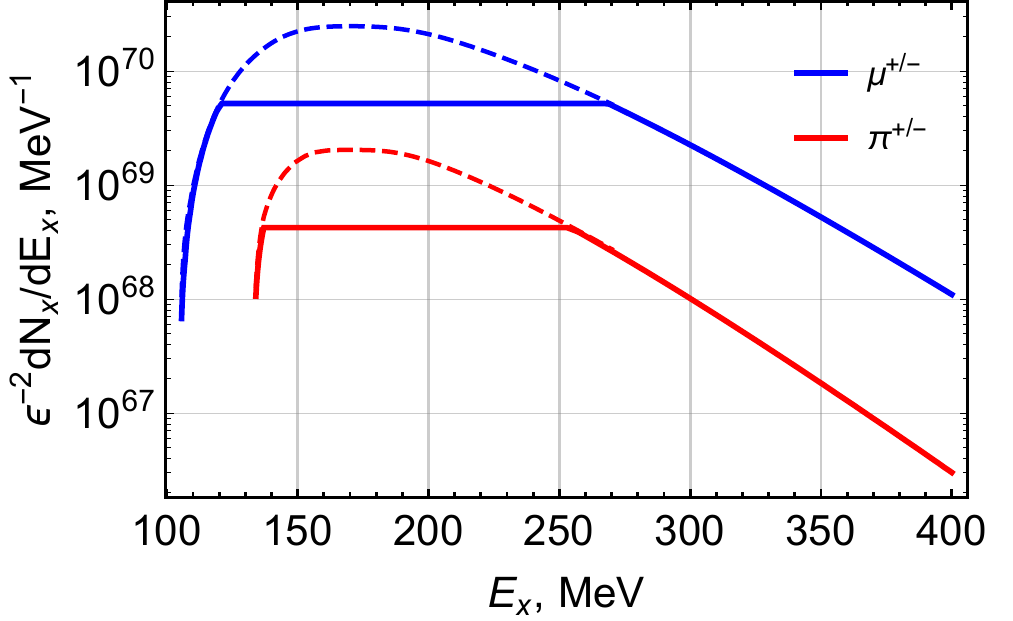}
    \caption{Energy spectra of pions and muons from dark photon decay with mass $m_{A'} = 340$~MeV (solid lines). The smaller quantity of pions results from a lower branching ratio for this decay channel. The distinction between dashed and solid lines is due to gravitational trapping.}
\label{fig:Spectra_of_muons_pions}
\end{figure}

An example of the resulting spectra of the pions and muons is given in Fig.\ref{fig:Spectra_of_muons_pions}
The suppression of the pion spectra is due to the small BR for the presented mass (BR$_{A'\rightarrow \pi^+\pi^-} \approx 3.7\%$ for $m_{A'} = 340$ MeV). For higher masses, the importance of pions will increase. The distinctive shape of the "flat" spectra is due to gravitational trapping, implemented as a low-energy cut-off for the spectra of dark photons. 

\section{Muons and pions inside the supernova}
\label{sec:propagation_mu_pi}

Muons and charged pions participate in the EM  interactions with the SN plasma -- mostly with electrons, protons, and photons. The temperature outside the SN core is significantly lower than the mass of pion or muon.
Therefore, if such interactions occur at a rate higher than the decay of the corresponding particles, they will cool down, losing their momenta and becoming non-relatistic.
We will call such a process \textit{thermalization}, implying that it is only a kinetic energy equilibration with plasma.

We start our analysis with the case of muons.
Three main thermalization channels are
\begin{enumerate}
    \item Scattering off photons $\mu^\pm + \gamma \rightarrow \mu^\pm + \gamma$
    \item Scattering off free electrons in plasma\footnote{the abundance of positrons at $T\ll 1$~MeV is negligible.}  $\mu^\pm + e^-\rightarrow \mu^\pm + e^-$
    \item Scattering on free protons\footnote{At large radii, there are also He nuclei, but as 
 we will see, this process is subdominant} $\mu^\pm + p \rightarrow \mu^\pm + p$
\end{enumerate}
\subsection{Scattering of muons on free electrons.} 
The stopping power and energy loss of $E_\mu\sim 100$~MeV muons in regular matter is dominated by the ionization processes \cite{Groom:2001kq}, which in the case of muon kinetic energy $T_\mu \sim \si{MeV}$ is equivalent to scattering on free electrons.  We can expect, that in the case of the SN plasma, the result will be similar.

Naively, the dimensional estimate for the cross-section would be similar to the Thomson scattering  $\sigma_T \sim \alpha^2/E^2$ with typical energy in reaction $E\sim m_\mu$, but instead, due to small electron mass and, as a result,  small transferred momentum, this cross-section will be significantly higher as well as energy-loss rate - see App \ref{App:E_Mu_therm}.

The temperature $T_{e, th}$, where such thermalization occurs can be estimated as 
 \begin{equation}
    N_{e, th}^{-1}\cdot \sigma_{\mu e} \cdot n_e (T_{e, th}) = \Gamma_{\dec}^{\mu}
 \end{equation}
 here, $ N_{e, th}$ - number of reactions needed to thermaliza muon, $\sigma_{e\mu}$ - electron-muon scattering cross-section, $n_e$ - electron number density. Such an estimate will lead to the thermalization temperature
$$T_{e, th} \sim 10^{-5}~\text{MeV}$$
 corresponding to radius $R \gtrsim 10^5-10^6$ km. As will be seen from the results, such a large radius would mean that the major part of the parameter space of the dark photons' parameters that are potentially interesting for us, would correspond to the thermalized muons.
 
\subsection{Muon-photon scattering} 
In a limit, where the energy of the photon is small $\omega \sim T \ll m_\mu$, the Thomson cross-section is applicable with the rate:
 \begin{equation}
    \Gamma_{\mu\gamma}(T) = \sigma^{\mu}_{T}\cdot n_\gamma (T), ~~~ \sigma^{\mu}_{T}=\left(\frac{m_e}{m_\mu}\right)^2~0.665~\text{b},
 \end{equation}
 $$n_\gamma(T) = 2\frac{\zeta(3)}{\pi^2}T^3 $$
 here $\sigma^{\mu}_{T}$ - Thomson scattering cross-section for muons. The energy carried by photons after scattering is
 \begin{equation}
     \omega' \approx \gamma^2 \omega (1 + \beta \cos\theta')(1 - \beta \cos\theta)
 \end{equation}
 where $\gamma,\beta$ - are Lorentz factors for incoming muon, $\theta, \theta'$ - angular directions before and after scattering, $\omega, \omega' $ - energy of the photon before and after scattering. In our scenario, the energy of the muon will not be ultra-relativistic and $\gamma \sim \mathcal{O}(1)$. To thermalize the muon $N_{\gamma, th} \sim \frac{E_\mu}{\gamma^2\cdot \braket{\omega}}$ reactions are needed. So, using the similar estimate as for the case of the electron scattering: 
 \begin{equation}
     N_{\gamma, th}^{-1} \sigma^{\mu}_{T} \cdot n_\gamma (T_{\gamma, th}) = \Gamma_{\dec}^{\mu}
 \end{equation}
  will give a typical value of thermalization temperature $$T_{\gamma, th} \approx 0.03 \text{MeV}$$  which corresponds to $R \sim 10^3$ km. Since this radius is much smaller than in the case of electrons,

\subsection{Scattering on protons}
The calculations for this process in the approximation of the point-like proton can be done with the replacement of the electron mass with the proton mass in the $ e - \mu$ scattering. As a result, it can be seen (see \ref{App:E_Mu_therm}), that due to the large proton mass, the rate of the energy loss due to muon-proton scattering is significantly lower, than in the case of electron-muon scenario.

Therefore, the $e +\mu \to e + \mu$ process will be dominant in the thermalization of muons. If the decay of the dark photon occurs at a higher temperature, the resulting muon spectra will become non-relativistic, and for further calculations, they are considered to be at rest at the moment of decay. On the other hand, if the temperature, where the decay of the DP occurs is lower, $\mu$'s spectra will still be given by Eqn.\eqref{eq:2-body_master_mu}.

The above-written can be similarly applied for pions with a replacement of muon mass with the pion mass and corresponding lifetimes. Besides, they can participate in strong interactions in elastic pion-proton or pion-neutron scatterings. But pions, as it will be seen, will give a subdominant contribution to the final result.
 
As was mentioned previously, to calculate the production of the dark photon the toy model of the uniform tsphere is used. It doesn't have any information about the temperature profile of the media outside the core. But to take into account the thermalization and its potential effect no simple model can be used. Therefore, a model of Electron-Capture Supernovae -  the 8.8  $M_\odot$  spherical symmetry simulation SN \cite{2010PhRvL.104y1101H,Garchinv_archive} has been used to obtain the radial temperature profiles. The examples of such hydrodynamical profiles are given in App \ref{App:Sn_profiles}.

\section{Neutrinos from dark photon decays}
\label{sec:DP_secondary_neutrinos}

In this Section we will analyze the spectra and flavor composition of the neutrinos, produced from dark photon decays. The neutrinos are produced in the decay of secondary particles (Section~\ref{sec:spectra_at_production}), change their flavor composition while propagating (Section~\ref{sec:oscillation}) and, finally, get detected at terrestial detectors (Section~\ref{sec:detection}).

\subsection{Spectra of secondary neutrinos at production}
\label{sec:spectra_at_production}

Muons decay primarily via
\begin{equation}
    \mu^- \rightarrow e^- +\bar\nu_e + \nu_\mu
    \label{eq:mu_decay_channel}
\end{equation}
(plus a CP-conjugated process).
This decay produces neutrinos of both muon and electron flavors. 
In the muon's rest frame, the distribution of these neutrinos in energies and
direction is given by:
\begin{equation}
    \frac{d^2N}{dE_\nu d\cos\theta} \equiv f_I(E_\nu)=\frac{256}{54}\frac{E_\nu^2}{m_\mu^3}\left(3-4\frac{E_\nu}{m_\mu}\right)
    \label{eq:neutrinos_3_body}
\end{equation}
(for $\bar\nu_\mu$ and $\nu_e$ in the case of CP-conjugated process)
\begin{equation}
    f_{II}(E_\nu)=48\frac{E_\nu^2}{m_\mu^3}\left(1-2\frac{E_\nu}{m_\mu}\right)
        \label{eq:antineutrinos_3_body}
\end{equation}
(for $\bar\nu_e$ and $\nu_\mu$ in the case of CP-conjugated process).
This leads to the following expression for the neutrino spectra in the lab frame, c.f.\ \cite{Mastrototaro:2019vug}:
\begin{equation}
    f_{\text{3-body}}(E) = \int dE_N~d\cos\theta \, \frac{dN_\mu}{dE_\mu} f_{a}\left(\frac{E}{\gamma(1+\beta \cos\theta)}\right)
    \label{eq:3-body_master}
\end{equation}
here $\beta = p_\mu/E_\mu,\gamma = E_\mu/m_\mu$ are the Lorenz factors for muons, $f_a$ - spectra of secondary neutrinos/antineutrinos Eqs.~\eqref{eq:neutrinos_3_body}--\eqref{eq:antineutrinos_3_body}, the subscript "a" stays for $I, II$. 

Next, consider the pion decay. 
The dominant decay channel for pions is \cite{ParticleDataGroup:2020ssz}:
\begin{equation}
    \pi^- \rightarrow \mu^- + \bar\nu_{\mu}
\end{equation}
and other channels can be ignored due to the large helicity suppression.
For this 2-body decay, the expression for the spectra of finalneutrinos is almost the same as in \eqref{eq:2-body_master_mu}. 
The only difference is the negligible masses of neutrinos, for which there is no upper limit in the integration:
 \begin{equation}
    f_{\nu}(E) = \frac{m_\pi}{2\bar{E_\nu}}\int_{\text{E}_{\text{min},\nu}}\frac{d E_\pi}{p_\pi}\frac{dN_\pi}{dE_\pi}
      \label{eq:2-body_master_pion_decay}
\end{equation}
 where $\bar{E_\nu} = \frac{m_\pi^2-m_\mu^2}{2m_\pi}$, and lower bound $\text{E}_{\text{min},\nu}$ is a solution for 
 \begin{equation}
    m_\pi \frac{E}{\bar{E_\nu}} = \text{E}_{\text{min},\nu} + \sqrt{\text{E}_{\text{min},\nu}^2-m_\pi^2}
\end{equation}

\subsubsection{Thermalization of muons and pions.} 
\label{sec:thermalization}
\begin{figure*}
    \centering
    \includegraphics[width=0.5\textwidth]{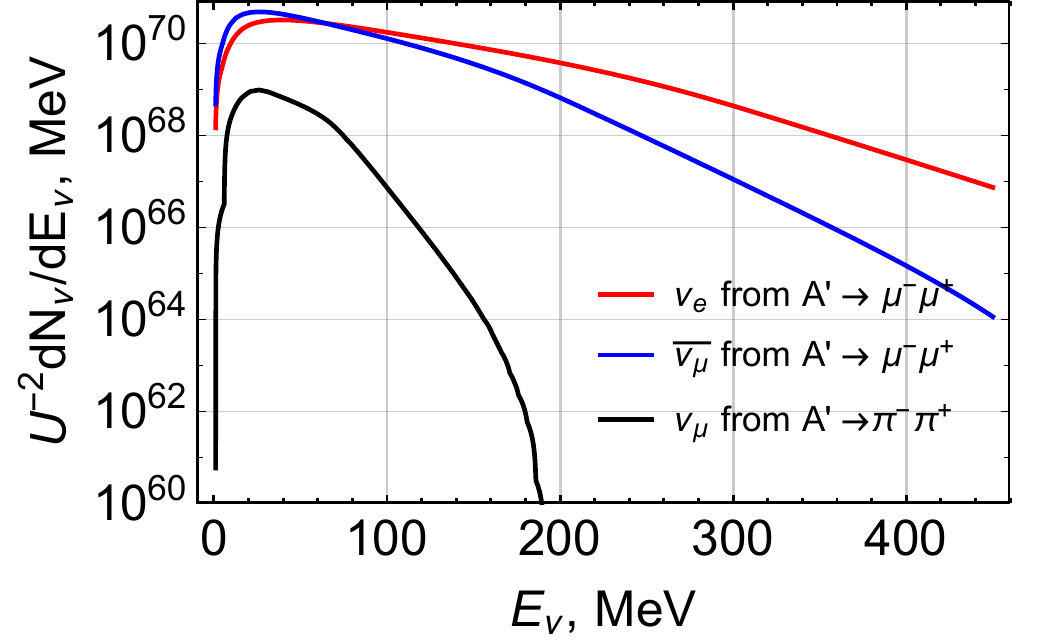}~ \includegraphics[width=0.5\textwidth]{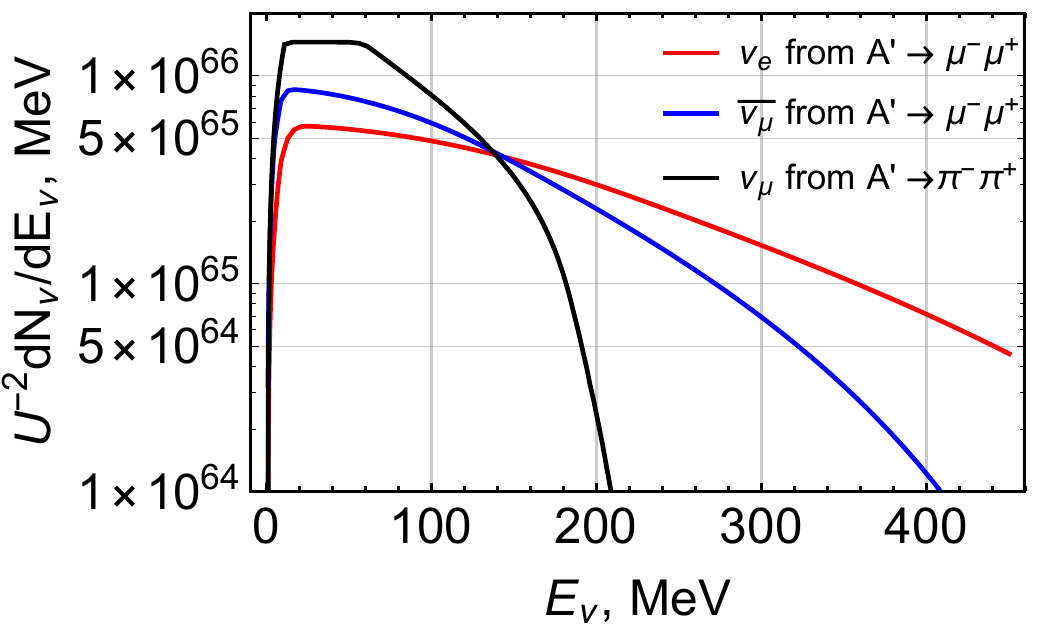}
    \caption{Spectra of secondary neutrinos of different flavors from the dark photon decays for two masses: $m_{A'} = 300$ MeV (left plot) and $m_{A'} = 600$ MeV (right plot) in a scenario, where muons and pions do not lose their energy \textit{e.g. non-thermalized}. Neutrinos from the pion channel have more narrow spectra due to their definite energy in the 2-particle decay process.}
    \label{fig:spectra_of_neutrinos_at_decay}
\end{figure*}

In the case of  $\mu$ and $\pi$ losing their energy during thermalization and becoming non-relativistic, the results for the neutrino spectra can be significantly simplified. In such a scenario the spectra  of neutrinos for each channel are proportional to the total number density of dark photons produced and the corresponding decay channel and are given by:
\begin{itemize}
    \item Decay into a pair of pions leads to a sharp peak  in $\nu_\mu$ spectrum 
    \begin{equation}
        \frac{dN_{\nu_\mu}}{dE} = \text{Br}_{A\rightarrow \pi^+\pi^-} N_{A'} \delta(E-\bar{E_\nu}), ~~~ \bar{E_\nu} \approx 29.8 \text{ MeV}
    \end{equation}
    \item Decay into a pair of muons leads to a spectrum of electron and muon flavor neutrinos, proportional to previously mentioned $f_{I, II}$
    \begin{equation}
        \frac{dN_{\nu_{\mu}}}{dE} = 2 \text{Br}_{A\rightarrow \mu^+\mu^-} N_{A'} f_{I}
    \end{equation}
    \begin{equation}
        \frac{dN_{\bar\nu_{e}}}{dE} = 2 \text{Br}_{A\rightarrow \mu^+\mu^-} N_{A'} f_{II}
    \end{equation}
\end{itemize}
Where the $N_{A'}$ - is the number of produced dark photons, factor $2$ appears from integration over $d\text{cos}\theta$ in the isotropic case. The examples of the spectra of neutrinos appearing in non-thermalized scenarios are shown in Fig.\ref{fig:spectra_of_neutrinos_at_decay}.
For smaller masses, the dominant channel is decay into a pair of muons while for the higher mass, decay into pion contributes more to the lower-energy part of neutrino spectra. 

\subsection{Propagation of neutrinos} 
\label{sec:oscillation}
Neutrinos, produced in decays will propagate freely outside of the SN media towards Earth. Due to oscillations, their flavor composition will change. Those could be regular vacuum oscillations ~\cite{Giganti:2017fhf} or matter-induced resonant conversions like the MSW effect \cite{Mikheev:1986gs,Wolfenstein:1977ue}. The latter process occurs inside the SN media at radii  $R_{\text{MSW}}\lesssim 10^5$ km \cite{Mirizzi:2015eza}). Neutrinos, that travel through the less dense environment, experience only vacuum oscillations.

\begin{figure}[!t]
    \centering
    \includegraphics[width=0.5\textwidth]{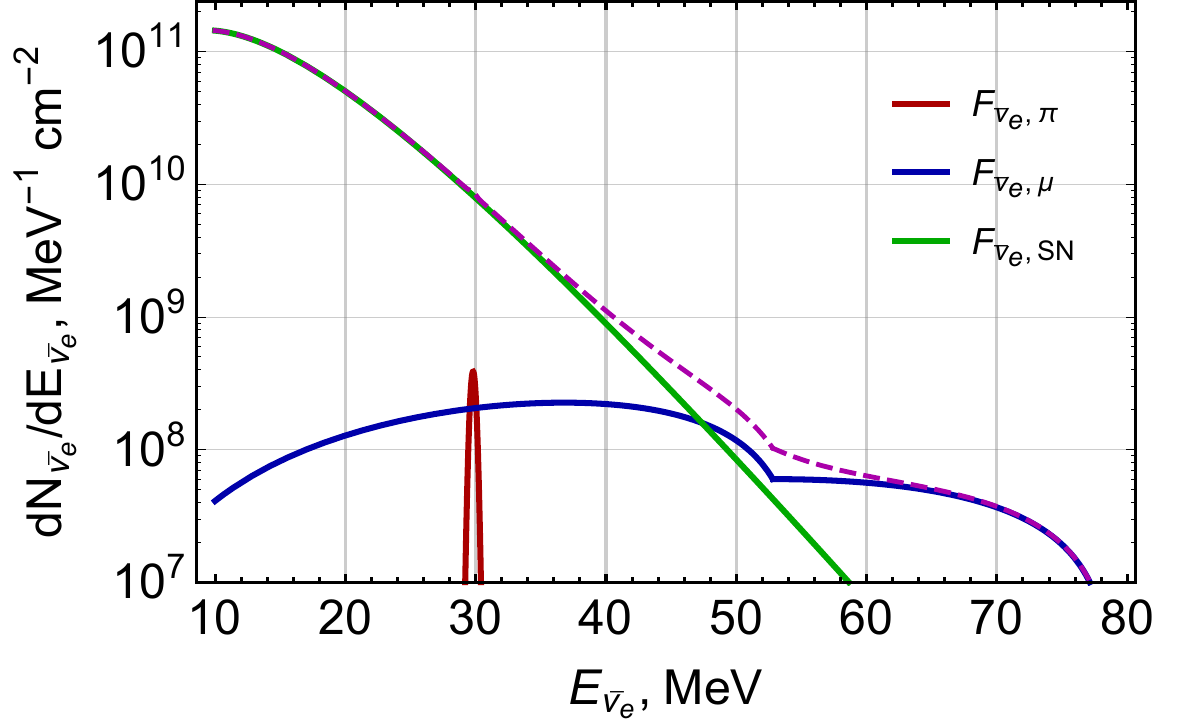}
    \caption{Flux of $\bar\nu_e$, that can be observed on Earth originating from muons and pions decay channels, compared to a typical spectrum of SN anti-neutrinos, approximated by Eqn.~\protect\eqref{eq:fit_SN_neutrinos}.
    Parameters of the dark photons are $m_{A'} = \SI{340}{MeV}$, $U^2 = 10^{-18}$. The blue line corresponds to anti-neutrinos, produced in muon decays, while the red line corresponds to the anti-neutrinos from the pion decays. 
    The contribution to the total spectra from the secondary neutrinos becomes dominant at energy $E_{\bar\nu_e} \gtrsim \SI{50}{MeV}$. 
    The contribution from non-thermalized muons (c.f.Section~\protect\ref{sec:thermalization}) is ignored. 
    The dashed line corresponds to the total neutrino spectra.}
    \label{fig:spectra_at_Earth}
\end{figure}

The case of vacuum oscillations is independent of the neutrino mass hierarchy. The flux of resulting neutrinos of electron  flavor after oscillation is given by
\begin{equation}
    F_{\bar\nu_e, \text{osc}} = P^{\text{vac}}_{ee} F_{\bar\nu_e}^0 + P^{\text{vac}}_{e\mu} F_{\bar\nu_\mu}^0 
\end{equation}
where $P^{\text{vac}}_{ee}$ - survival probability for electron neutrino, $P^{\text{vac}}_{e\mu}$ - oscillation probability from a muon flavor, $F_{\bar\nu_\alpha}^0$ for $\alpha = \{e,\mu,\}$ - spectra of the secondary neutrinos. Tau flavor is omitted as in DP decays it will not be produced. The standard expression for vacuum oscillating, time-averaged probabilities, for the case of vanishing complex phases in the PMNS matrix (see Ref.\cite{Mirizzi:2015eza}), is:
\begin{equation}
    P_{\nu_\alpha \to \nu_\beta} = \delta_{\alpha\beta} - 2 \sum_{i>j} U_{\alpha i}U_{\alpha j} U_{\beta i} U_{\beta j}
\end{equation} 
here $U_{\alpha \beta}$ - components of PMNS matrix (see Ref.\cite{Giganti:2017fhf}) with mixing angles taken as best-fit values from Ref.\cite{Capozzi:2013csa}:
\begin{eqnarray}
\sin^2\theta_{12} = 0.308 \nonumber\\
\sin^2\theta_{13} = 0.023\nonumber\\
\sin^2\theta_{23} = 0.437\nonumber
\end{eqnarray}
In the case of MSW effect, the resulting $\bar\nu_e$ spectrum  is given by (c.f.~\cite{Mirizzi:2015eza}):
\begin{equation}
\label{eq:RegII}
    F_{\bar\nu_e} = \bar P^{\text{MSW}}_{ee}F_{\bar\nu_e}^0 + [1-\bar P^{\text{MSW}}_{ee}]F_{\bar\nu_x}^0
\end{equation}
where $F_{\bar\nu_e}^0$ --  flux of the secondary $\bar\nu_e$,  $F_{\bar\nu_x}^0$ -- flux of  muon or tau-flavor secondary neutrinos. 
Survival probability $\bar P^{\text{MSW}}_{ee}$ depends on the hierarchy of neutrino masses: $\bar P^{\text{MSW}}_{ee} = \cos^2\theta_{12} \approx 0.692$ for NH and $\bar P^{\text{MSW}}_{ee} = 0$ for IH. 
The example of the flux of secondary electron anti-neutrino arriving at Earth from the SN at distance $R_{\text{SN}} = 10$ kpc after oscillations taken into account is shown in Fig.\ref{fig:spectra_at_Earth}.

Our resulting (anti)neutrino spectra should be compared to the SN neutrino spectra.
The latter can be described by the following fitting formula, \cite{Tamborra:2012ac}:
\begin{widetext}
  \begin{equation}
    \frac{dN_{\nu}}{dE} = L_{\nu} \frac{(1+\alpha)^{1+\alpha}}{\Gamma (1+\alpha)\braket{E_{\nu}}^2}\left(\frac{E}{\braket{E_{\nu}}}\right)^\alpha \text{Exp}\left[-(1+\alpha)\frac{E}{\braket{E_{\nu}}}\right]
    \label{eq:fit_SN_neutrinos}
  \end{equation}
\end{widetext}
Here $L_{\nu} = 0.5\cdot 10^{53}$ erg is the total emission energy in one neutrino flavor, it is assumed to be the same for all flavors of neutrinos/anti-neutrinos; $\braket{E_{\nu}}$ are $\braket{E_\nu^2}$ is the average energy (energy squared) of the neutrinos; $\alpha$ is the ``pinching'' parameter, that indicates how do the spectra differ from the Fermi-Dirac distribution. 
It is related to the average energy and average squared energy as: 
\begin{equation}
  \label{eq:Enu_spectrum}
    \frac{\braket{E_{\nu}^2}}{\braket{E_{\nu}}^2} = \frac{2+\alpha}{1+\alpha}
\end{equation}
The numeric values used to describe the SN neutrino spectrum are given in Table~\ref{tab:fitting_aprameters}.
Finally, $\Gamma(\alpha)$ is the Euler's Gamma function.

\subsection{Detection of secondary neutrinos}
\label{sec:detection}

\begin{table}[!t]
\begin{tabular}{ |c|c|c| } 
 \hline
  & $\alpha$ & $\braket{E_{\nu}}$, $[\text{MeV}]$ \\ 
  \hline
 $\bar\nu_e$ & 2.78 & 12.69 \\ 
 $\nu_e$ & 2.9 &  10.14 \\ 
  $\nu_x/\bar\nu_x$ & 2.39 & 12.89 \\
  \hline
 \end{tabular}
 \caption{Fitting parameters for the SN neutrinos spectra, Eqs.~(\ref{eq:fit_SN_neutrinos}--\ref{eq:Enu_spectrum}).}
 \label{tab:fitting_aprameters}
\end{table}
\begin{figure*}[t!]
    \centering
    \includegraphics[width=0.45 \textwidth]{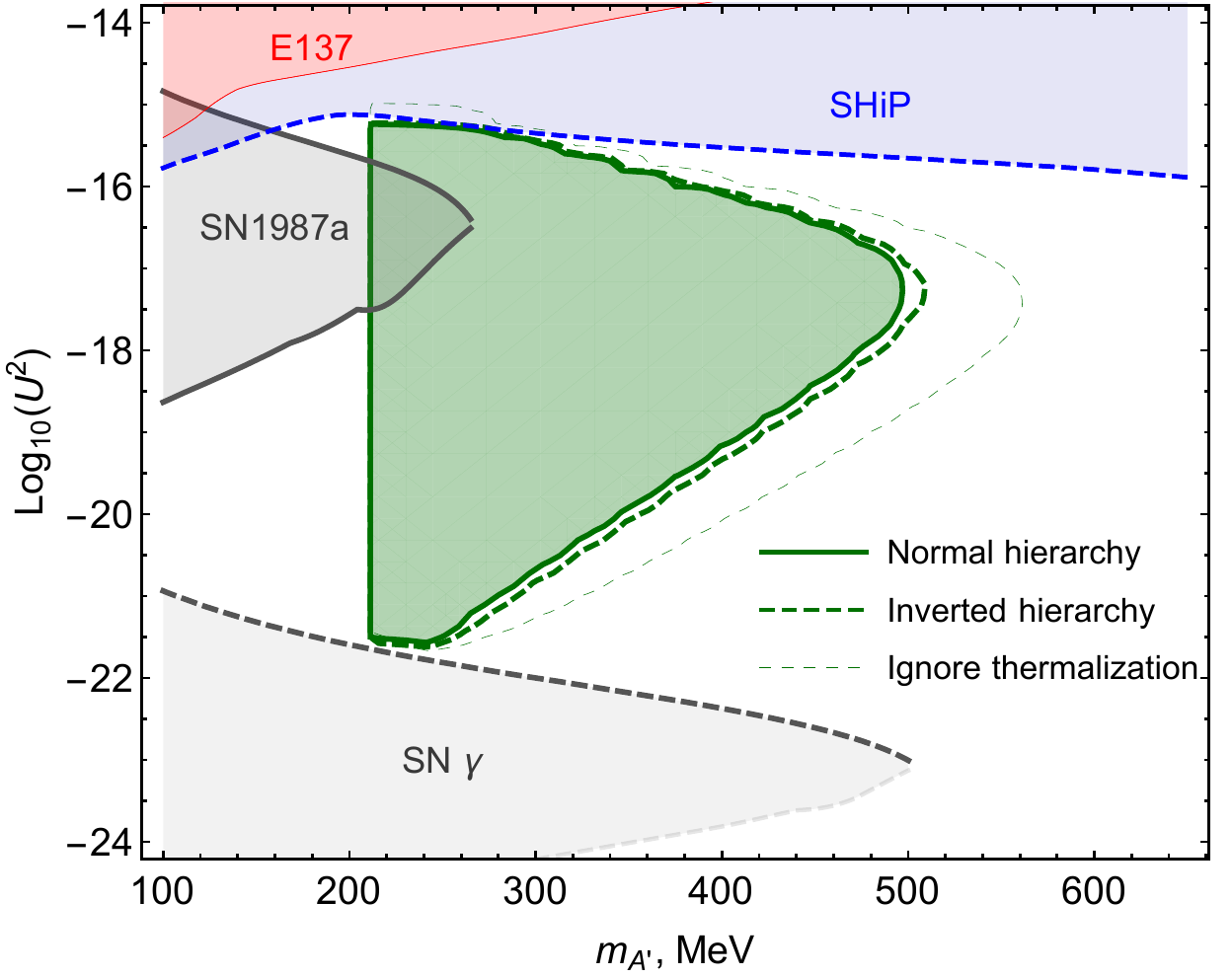}~
    \includegraphics[width=0.45 \textwidth]{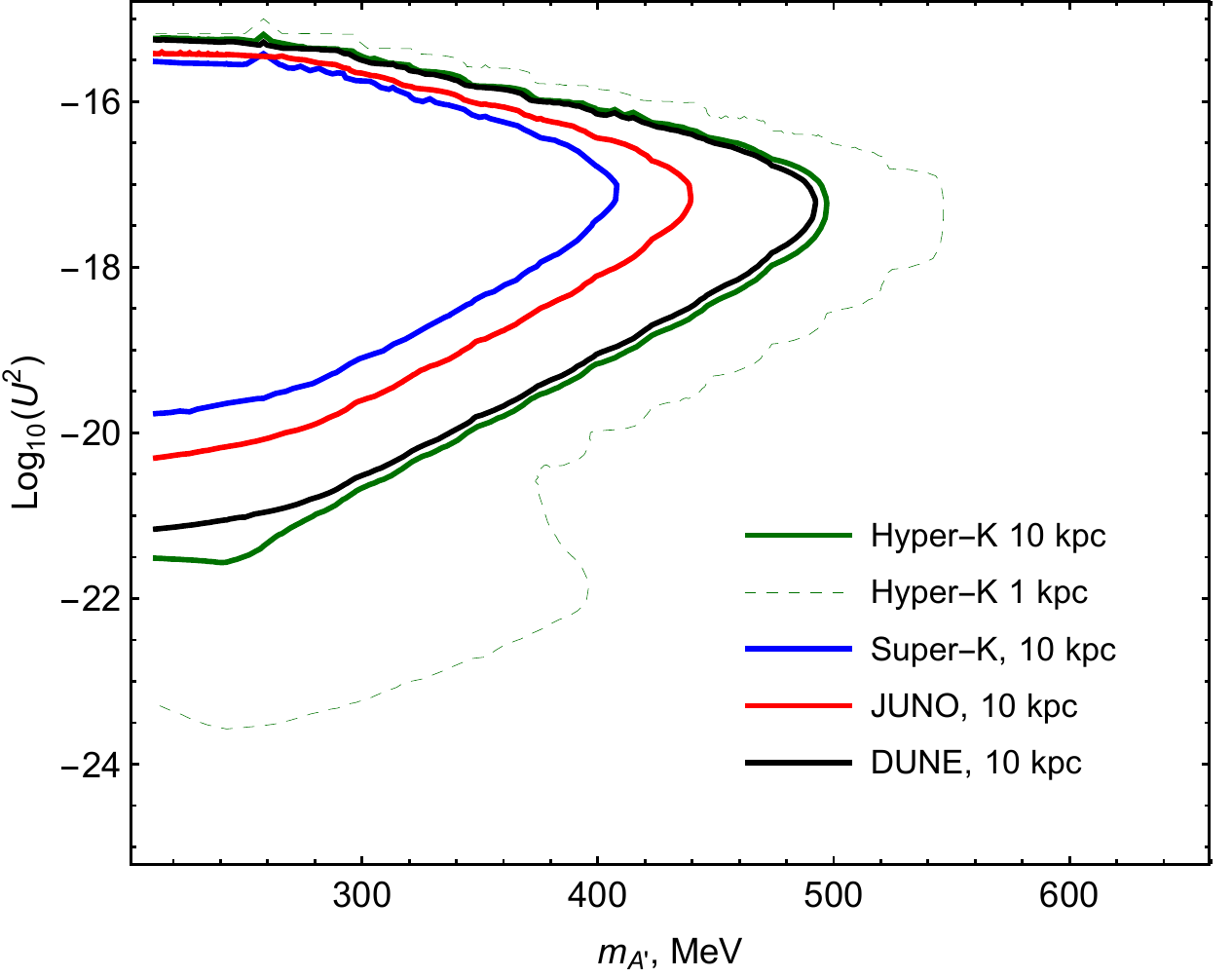} 
    \caption{\textbf{Constraints (95\% CL) on dark photon parameters based on the non-detection of high-energy neutrinos in the event of a supernova explosion.}
      \emph{Left panel:} 
      The green-shaded region marks the parameter space where at least three high-energy neutrino events are expected, assuming the dark photon's existence and normal hiererchy of neutrino masses.
      The difference between the solid and thin dashed green lines shows the muon thermalization effect, with the latter assuming no thermalization (see text for details). Thick green dashed line corresponds for inverted hierarchy.
      The red shaded area indicates constraints from SLAC's E137 experiment~\protect\cite{Batell:2014mga}.
      Gray areas show dark photon limits from SN1987A: darker gray is based on the energy loss arguments \cite{Chang:2016ntp}, while the lighter gray comes from the absence of electromagnetic signal fluctuations \cite{Kazanas:2014mca}.
      The reach of the future SHiP experiment is also shown for comparison \cite{SHiP:2018xqw}.
      All SN results assume the Hyper-Kamiokande detector, a 10~kpc supernova distance, and best-fit neutrino oscillation parameters for the normal ordering scenario.
      \emph{Right panel:} dependence of the sensitivity region on different assumptions (distance, mass ordering, detector type), as indicated in the legend, normal mass ordering is assumed here.}
    
    \label{fig:Main_Result}
\end{figure*}
Neutrinos from supernova explosions and from dark photon decays will be detected by terrestrial experiments.
The Hyper-Kamiokande, an underground water Cherenkov detector with around 220~kton of ultra-pure water as its detection volume, is the most promising in terms of expected event numbers. It is anticipated to detect approximately $10^5$ events from the next supernova in the galactic center. We will base most of our estimates on this detector.
Two other neutrino detectors of interest are DUNE and JUNO.\footnote{Their characteristics, relevant for our analysis, are summarized in Table~\protect\ref{tab:expected_events}.} The former is the liquid argon time-projection chamber (LArTPC).
It is designed to have a superior energy resolution compared to water or organic-based scintillators. The latter is a liquid scintillator detector. Both experiments are expected to detect around $N_{SN} \simeq (5- 10) \times 10^3 $ events which include not only electron flavor neutrinos but also neutral current events from elastic scattering of neutrinos of other flavors. The dominant process for each experiment is inverse beta decay (IBD).
\begin{itemize}
    \item For  Super-K, Hyper-K and JUNO -- the capture of electron anti-neutrino via $$ \bar\nu_e + p \rightarrow e^+ + n\,.$$
    \item For DUNE -- the capture of electron neutrino via $$\nu_e +^{40}\text{Ar}\rightarrow e^- + ^{40}K^*\,.$$
\end{itemize}

The estimated number of events per electron (positron) recoil energy, $T_e$, is given by:
\begin{widetext}
\begin{equation}
    \label{eq:number_detected_IBD}
    \frac{d N_e}{d T_e} = \frac{\epsilon\, N_c}{4\pi R_{SN}^2}\int\limits_{E_{\rm break}}^{\infty}  d E \,\frac{d N_{\bar\nu}(E)}{dE} \frac{d \sigma_{\text{\sc ibd}}(E, T_e)}{dT_e}\,.
\end{equation}
\end{widetext}
Here $\epsilon$ is the detector efficiency; $N_c$ is the number of targets in the detector\footnote{Protons in the case of Hyper-K and JUNO and Argon in the case of DUNE} are parameters of the experiment, see Table~\ref{tab:expected_events}.
The differential cross-section for inverse beta-decay $d\sigma_{\text{\sc ibd}}/dT_e$ depends both on both (anti)-neutrino energy and on the recoil energy $T_e$. 
The integration over (anti)neutrino energies starts from $E_{\rm break}$ -- the energy above which the contribution of the secondary neutrinos becomes dominant over the SN neutrinos and the observed spectrum exhibits a break (c.f.\ Fig.~\ref{fig:spectra_at_Earth} where the $E_{\rm break} \approx \SI{52}{MeV}$).
The $R_{SN}$ is the distance to the SN.  Finally, $\frac{d N_\nu(E_\nu)}{dE_\nu}$
is the spectrum of (anti)neutrinos, as computed in
Section~\ref{sec:DP_secondary_neutrinos} above.
The total number of events will be given by
\begin{equation}
  \label{eq:N_e_total}
  N_{\text{total}} =\int\limits_{E_{\rm break}}^{\infty} dT_e\, \frac{dN_e}{dT_e}\,.  
\end{equation}

\begin{table}[!t]
  \begin{tabular}{l|c|c|c|c}
    \toprule
        Parameters & Hyper-K & Super-K & JUNO & DUNE \\
      \midrule
      Fiducial volume & $\SI{187}{kton}$ &  $\SI{22.5}{kton}$ & $\SI{20}{kton}$ & $\SI{40}{kton}$\\
      Efficiency, $\epsilon$ & 0.67 & 0.67 & 0.8 & 0.86 \\
      Number of targets, $N_c$ & 2.5 $\cdot 10^{34} $ & 1.5 $\cdot 10^{33} $  & 1.2 $\cdot
                                                             10^{33}$ & 6
                                                                        $\cdot
                                                                        10^{32}
                                                                        $  \\
    \midrule
      Refs. & \cite{Hyper-Kamiokande:2016srs} & \cite{Super-Kamiokande:2008mmn} & \cite{JUNO:2021vlw} & \cite{DUNE:2021tad}\\
        \bottomrule
    \end{tabular}
    \caption{Characteristics of the neutrino detectors, needed to estimate the number of detected events (see Eq.~\protect~(\ref{eq:number_detected_IBD}--\ref{eq:N_e_total})): the
      detector efficiency $\epsilon$ and the number of targets, $N_c$.}
    \label{tab:expected_events}
\end{table}

\section{Results}
\label{sec:results}
We express our results as the region in the dark photon parameter space that can be explored should a supernova explosion be registered by terrestrial neutrino detectors.
Figure~\ref{fig:Main_Result} (left panel) shows the sensitivity of the Hyper-Kamiokande experiment.
The green shaded area is the region of the parameters space, where one expects detection of $N_{\text{total}} \ge 3$ high-energy neutrinos.
As the relevant range of neutrino energies ($\SIrange{50}{100}{MeV}$) is
essentially background free,\footnote{At these energies, the influence of supernova neutrinos becomes negligible. Consequently, atmospheric neutrinos are the sole source of background, with an occurrence rate of a few dozen events annually \cite{Akita:2022etk}. In contrast, we anticipate a sizeable number of events occurring within a timeframe of approximately 10~sec.} such an estimate corresponds to the 95\% confidence level.
The upper boundary of the region is set by the trapping effect.
For higher coupling constants, dark photons decay inside the supernova core, and their decay products thermalize without influencing the signal.
The vertical boundary on the left indicates the mass threshold \(m_{A'} \ge 2m_\mu\), below which DPs decay to \(e^-e^+\) pairs, which
could impact the supernova electromagnetic signal (c.f.\ \cite{Kazanas:2014mca}).

The neutrino signal of interest primarily arises from the decay of ``thermalized'' muons, which have dissipated their energy through electromagnetic interactions with the plasma, as elaborated in Section~\protect\ref{sec:propagation_mu_pi}.
The thermalization makes our findings independent of the temperature and density profiles outside the SN core.
Such profiles have never been measured and bear large theoretical uncertainties (see e.g.\ discussion in~\cite{Syvolap:2019dat}).
The fact that most of the muons have thermalized, makes our prediction dependent \emph{only} on the total number of produced dark photons inside the SN core, but not on the details of the supernovae' interior. 
Only a small ``bump'' at masses $m_{A'} \sim \SI{250}{MeV}$ and $U^2 \sim 10^{-22}-10^{-21}$ comes from the remaining ``non-thermalized'' muons (notice that this ``bump'' grows into a noticeable region for a nearby supernova at distance of 1~kpc, Figure~\ref{fig:Main_Result}, right panel).
Additionally, we examine the scenario where muons do not undergo thermalization (represented by the green dashed line).
This contour is largely model-dependent, serving to highlight the uncertainty involved in defining the boundary of the green region.
All the results on the left assume the 10 kpc distance to the SN and best-fit neutrino parameters for the normal mass ordering.

In the right panel of Figure~\ref{fig:Main_Result}, we demonstrate the sensitivity of our results to these assumptions, by varying the distance, and the type of the experiment and changing from normal to inverted mass ordering.
All limits are 95\% CL.

The left panel also includes a comparison of our results to both past and anticipated experiments. The results we have obtained occupy a region of the dark photon parameter space that has not been previously explored, nor is it expected to be covered in the near future. The gray shaded areas in the left panel represent other supernova constraints: those based on arguments regarding supernova energy loss \cite{Chang:2016ntp,Dent:2012mx,Kazanas:2014mca}, as well as constraints from the electromagnetic decays of DPs, which lead to fluctuations in the photon signal \cite{Kazanas:2014mca}. Additionally, we show the coverage by relevant particle physics experiments: the SLAC Beam Dump Experiment E137 \cite{Batell:2014mga}, and the expected sensitivities from the forthcoming SHiP experiment \cite{SHiP:2018xqw}.



\section{Conclusion and discussion}
\label{sec:conclusion}

Supernova interiors are considered potential birthplaces for hypothetical feebly interacting particles (axions, dark photons, sterile neutrinos, etc), which can have masses up to several hundred MeVs, see e.g.\ \cite{Antel:2023hkf}.
In the past, SN constraints have been mostly driven by the energy-loss arguments that aim to restrict the energy emission rate from the SN via the FIP channel to limit the potential shortening of active neutrino emission (see e.g.\ \cite{Raffelt:1996wa}).
However, the neutrino signal from a supernova contains more information (a form of the spectrum, duration of the burst, etc) that can also be used to provide constraints on FIP parameters.

In this study, we investigated the generation of dark photons in supernovae and followed their subsequent escape from the supernova core and decay into secondary particles (muons and pions). The latter subsequently decay into neutrinos. This sequence of decays, with neutrinos as tertiary particles, is a novel area of exploration in the literature. We show that these neutrinos manifest as a high-energy ``tail'' in the neutrino distribution, making them distinctly identifiable.
Furthermore, such neutrino events occur in a part of the spectrum that is largely free from any background neutrino events, thereby facilitating their detection.

Consequently, a supernova explosion within our Galaxy could enable the probing of dark photon parameters using detectors such as Hyper-Kamiokande, DUNE, and to a lesser extent, Super-Kamiokande and JUNO. For supernovae occurring near the Galactic Center, the derived constraints would mainly pertain to neutrinos from the decays of thermalized muons, providing constraints that are largely independent of the supernova's temperature and density profiles and thus are robust. The contribution of neutrinos from non-thermalized muons (or from pion decay products) is minor at distances equivalent to the Galactic Center but becomes significant for supernova explosions much closer than the Galactic Center, offering the potential to explore an even broader portion of the parameter space.
We demonstrate that such an analysis would allow us to explore the region of dark photon's \textcolor{blue}{masses up to $m_{A'} = 500$ MeV and values of mixing parameter $U^2 \sim 10^{-21} - 10^{-16}$ which makes our result complementary for the future SHiP experiment and the excessive gamma-emission constraints, covering previously unexplored region. The developed novel approach will allow for the study of other short-living FIP species, that can be produced in the supernova environment and subsequently decay into neutrinos.  In the case of a galactic center supernova, it can provide higher sensitivity to the parameters of the FIP and less model dependence than the scenario of energy-loss/energy-transfer constraint.  }

\onecolumngrid

\appendix

\section{Distribution function of the decay products}
\label{app:distribution_function}

 Sections~\ref{sec:Dp_decays} and \ref{sec:DP_secondary_neutrinos} of this paper use the formula for the transformation between the phase-space distribution function of the parent particles in laboratory frame and the corresponding distribution function of one of the decay products. 
Such a relation is given by the following integral (see e.g.\ \cite{Mastrototaro:2019vug,Akita:2022etk}):
\begin{equation}
    \label{eq:Lorentz_transform_df}
    \frac{dN_x}{dE} = \int d(\cos\theta) \int d E_p \frac1{\gamma (1 + \beta \cos\theta)} \frac{dN_p}{dE_p} 
    f_{\star} \left(E_\star , \cos\theta \right)
\end{equation}
Here $dN_p/dE_p$ is the distribution function of the parent particle in the laboratory frame, $\gamma$ is the Lorentz gamma factor for transforming between the rest and laboratory frames, with  
\begin{equation}
    \label{eq:gamma-beta}
    \gamma = \frac1{\sqrt{1 - \beta^2}},
\end{equation}
and $E_\star$ is the energy of the decay product in the rest frame of the parent particle, corresponding to the energy $E$ in the laboratory frame, such that 
\begin{equation}
    E = \gamma(1 + \beta \cos\theta) E_\star
    \label{eq:En_lorentz}
\end{equation}.
In the case of the two-body decays, the distribution function of the decay product $x$ does not depend on the angle $\theta$ and is given by the $\delta$-function:
\begin{equation}
    \label{eq:2body_df}
    f_{\star} \left(E_\star , \cos\theta \right)  = \frac 12 \delta(E_\star - \bar E)
\end{equation}
where $\bar E$ is the energy of the decay product $x$ in the rest frame of the parent particle. Using $E_\star$ as a variable instead of $\cos \theta$ 
\begin{equation}
    d\cos\theta = -\frac{1}{\beta \gamma}\frac{E}{E_\star^2}dE_\star
    \label{eq:change_of_var}
\end{equation}
we can remove the delta function and one integration resulting in Eqn~\ref{eq:2-body_master_mu}. Since the delta function does not always have a solution, there appears a minimum value of the energy of the parent particle that can produce a decay product with given energy $E$. It corresponds to the case $\cos\theta = 1$ and appears as a solution of
\begin{equation}
    E = \frac{E_p + \sqrt{E_p^2-m_p^2}}{m_p}\bar E
 \end{equation}
 with respect to $E_p$, where $m_p$ - is the mass of the parent particle.
 In the case of 3-body decay, we can not simplify the integration and have to take both integrals.

\section{Integrands for NP and PP processes}
\label{app:Integrands}
Following the \cite{Dent:2012mx}, introduce variables:
\begin{equation}
    x = \frac{E_{A}}{T}; ~~~ y = \frac{m_\pi^2}{m_N T}; ~~~ q = \frac{m_{A'}}{T}; ~~~ z = cos(\theta_{ij})
\end{equation}
where $E_{A}$ - the energy of the dark photon and $\theta_{ij}$ - scattering angle between nuclei in the center of the mass frame.
In those terms, the energy emission integral can be simplified (using the approximation $E_A, T \ll m_N$). PN-process integrand can be given as:

\begin{equation}
\mathcal{I}_{pn} = \sum_{j} C_j \mathcal{ I}_j
\end{equation}
where the components $\mathcal{I_j}$ and $C_j$ are:
\begin{eqnarray}
    I_1 &=& \frac{(u+v-2z\sqrt{uv})^3}{(u+v-2z\sqrt{uv}+y)^2}\\
    I_2 &=& \frac{(u+v-2z\sqrt{uv})(u+v+2z\sqrt{uv})^2}{(u+v+2z\sqrt{uv}+y)^2} \\
    I_3 &=& \frac{(u+v-2z\sqrt{uv})(-u^2-v^2+(6-z^2)uv)}{(u+v+y)^2-4z^2uv} \\
    I_4 &=& x\frac{-u^2-v^2+(6-4z^2)uv)}{(u+v+y)^2-4z^2uv}\frac{u-v}{u+v+y+2z\sqrt{uv}-\frac{T}{m_N}q^2}\\
    I_5 &=& x\frac{(u+v+2z\sqrt{uv})^2}{(u+v+2z\sqrt{uv}+y)^2}\frac{u-v}{u+v+y+2z\sqrt{uv}-\frac{T}{m_N}q^2}\\
    I_6 &=& x^2\frac{(u+v+2z\sqrt{uv})^3}{(u+v+2z\sqrt{uv}+y)^2}\frac{1}{(u+v+y+2z\sqrt{uv}-\frac{T}{m_N}q^2)^2} \\
    C_1 &=& 1\\
    C_2 &=& 4 \left[1+\frac{6x^2-4x(u-v)+2q^2}{(u+v)^2-4z^2uv}\right]\\
    C_3 &=& -2 \left[1+\frac{2x(u-v)}{-u^2-v^2+(6-4z^2)uv}\right]\\
    C_4 &=& -4\\
    C_5 &=& -16\\
    C_6 &=& 16
\end{eqnarray}
 In similar terms, the PP-integrand after the approximation and reduction will be:
\begin{equation}
    \mathcal{I}_{pp} = \frac{(u+v-2z\sqrt{uv})^3}{2 (u+v-2z\sqrt{uv}+y)^2}
\end{equation}

\section{Thermalization in $e\mu$ and $ep$ scatterings}
\label{App:E_Mu_therm}
\begin{figure}
    \centering
    \includegraphics[width=0.45\textwidth]{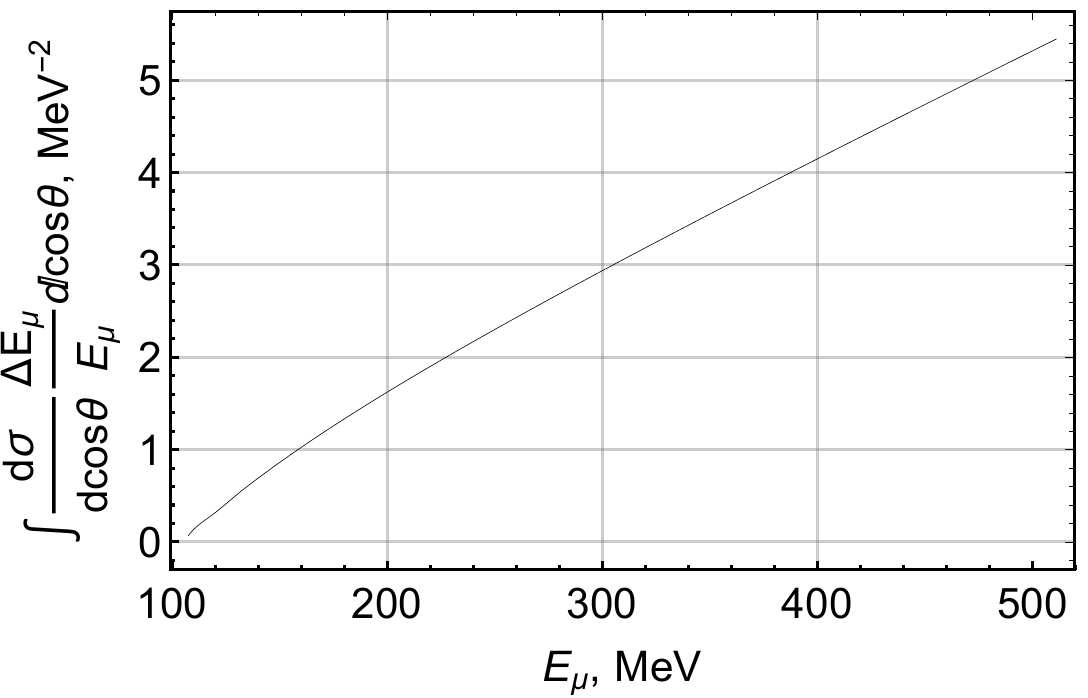}
    \caption{\textbf{Left panel:} Relative energy loss $\frac{\Delta E_\mu}{E_\mu}$ of muon in electron-muon scatterings averaged with differential cross-section. Typical muons from dark photons decays will lose $\approx 2-3 \%$ of their energy in each scattering, which is higher, than in the case of Thomson scattering where their energy loss at relevant SN radius with $T \sim 0.1$ MeV is an order of magnitude less. }
    \label{fig:muon_therm_CS}
\end{figure}

\begin{figure}
    \centering
    \includegraphics[width=0.45\textwidth]{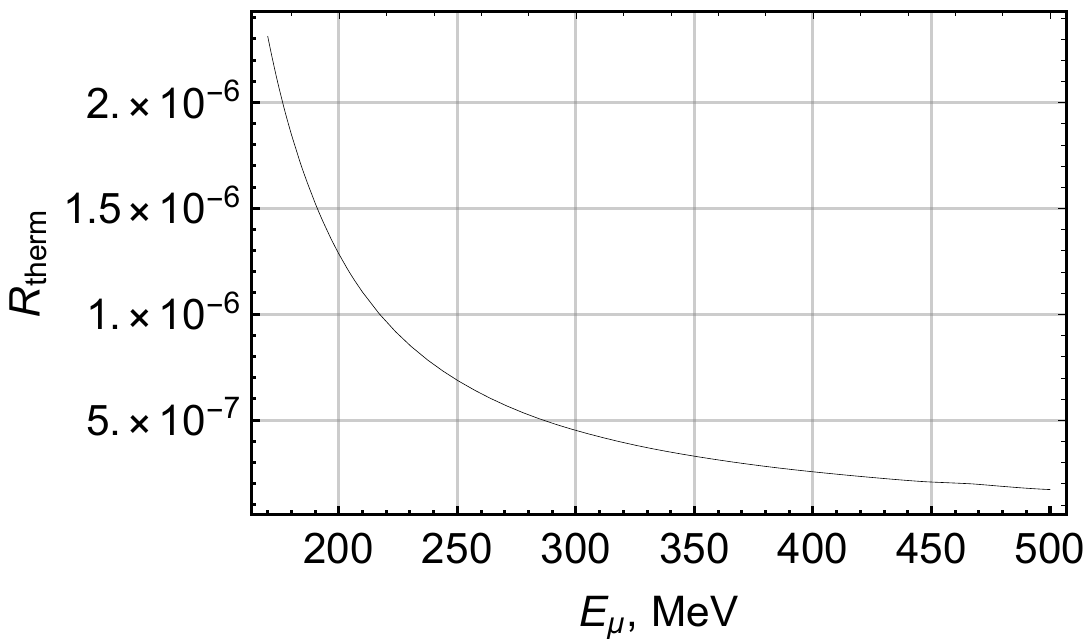}
    \caption{The ratio of muon energy loss rates in reactions with protons to the same process, but with electrons. The difference is at the level of several orders of magnitude, therefore we do not need to improve the calculations, considering the proton as a non-point-like particle as it will not change the outcome significantly and we can conclude that muon-electron interaction is the dominant process for the muon thermalization.}
    \label{fig:therm_ratio}
\end{figure}

The squared element for the electron muon scattering \cite{Schwartz:2014sze} is given by
\begin{equation}
    \frac{1}{4}\sum_{\text{spins}}|\mathcal{M}|^2 = \frac{2e^4}{t^2}\left[u^2 +s^2 +4t(m_e^2+m_\mu^2)-2(m_e^2+m_\mu^2)^2\right]
\end{equation}
where s,t, and u are Mandelstamm variables. The differential cross-section in the lab frame:
\begin{equation}
    \frac{d\sigma}{d\Omega} = \frac{1}{64\pi^2}\frac{1}{|\Vec{p_\mu}| m_e}\frac{1}{4}\sum_{\text{spins}}|\mathcal{M}|^2\frac{|\Vec{p'_\mu}|}{E_\mu + m_e - \frac{|\Vec{p_\mu}| E_{\mu}'}{|\Vec{p'_\mu}|}\cos \theta}
\end{equation}
where $\Vec{p_\mu},\Vec{p'_\mu}$ - 3-momenta of muon before and after the scattering, $E_{\mu},E_{\mu}'$ - energies of muon. Value of $|\Vec{p'_\mu}|$ is found from energy-momentum conservation:
\begin{equation}
    E_\mu+m_e = E_\mu' + \sqrt{|\Vec{p_\mu}|^2-2|\Vec{p_\mu}||\Vec{p'_\mu}|\cos\theta +|\Vec{p'_\mu}|^2 + me^2}
\end{equation}
the total cross-section requires regularization as it diverges at $\theta \to 0$. For simplicity of the estimates we use a relation between the impact parameter and the scattering angle in a classical case 
\begin{equation}
    b = \frac{\alpha}{2 E_\mu}\sqrt{\frac{1+\cos\theta}{1-\cos\theta}}
\end{equation}
and use the maximum value of the impact parameter as $b_{max} = n_e^{-1/3}$  where $n_e$ - is the number density of electrons. Since we are interested in the energy-loss rate, the actual cutoff will not change the result significantly.  The value of the averaged energy loss per reaction is given in Fig.\ref{fig:muon_therm_CS}
In the case of point-like protons, the mass of the electron in the above expression has to be replaced with the mass of a proton. The energy-loss rate on protons is significantly lower. the comparison between energy-loss on protons and electrons is presented in \ref{fig:therm_ratio}, where we plot a ratio:
$$R_{\text{therm}} = \frac{\sigma_{p\mu} \Delta E_\mu^{p\mu}}{\sigma_{e\mu} \Delta E_\mu^{e\mu}}$$
where indices $e\mu, p\mu$ correspond to the process with a given cross-section $\sigma$  and  energy loss $\Delta E$, averaged over angles. Since the number densities of protons and electrons are equal or very close in a major part of the SN due to electroneutrality, this ratio will also show the ratio between energy-loss rates in those two processes.

\section{Supernova simulation profiles}
\begin{figure*}[t!]
    \centering
    \includegraphics[width=0.32 \textwidth]{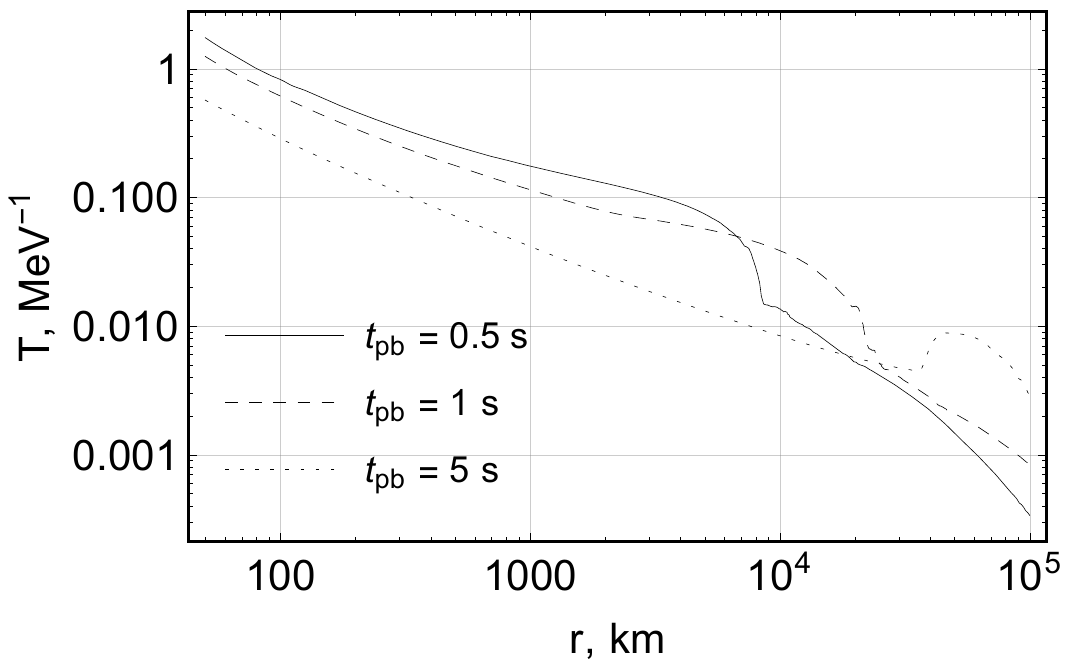}~
    \includegraphics[width=0.32 \textwidth]{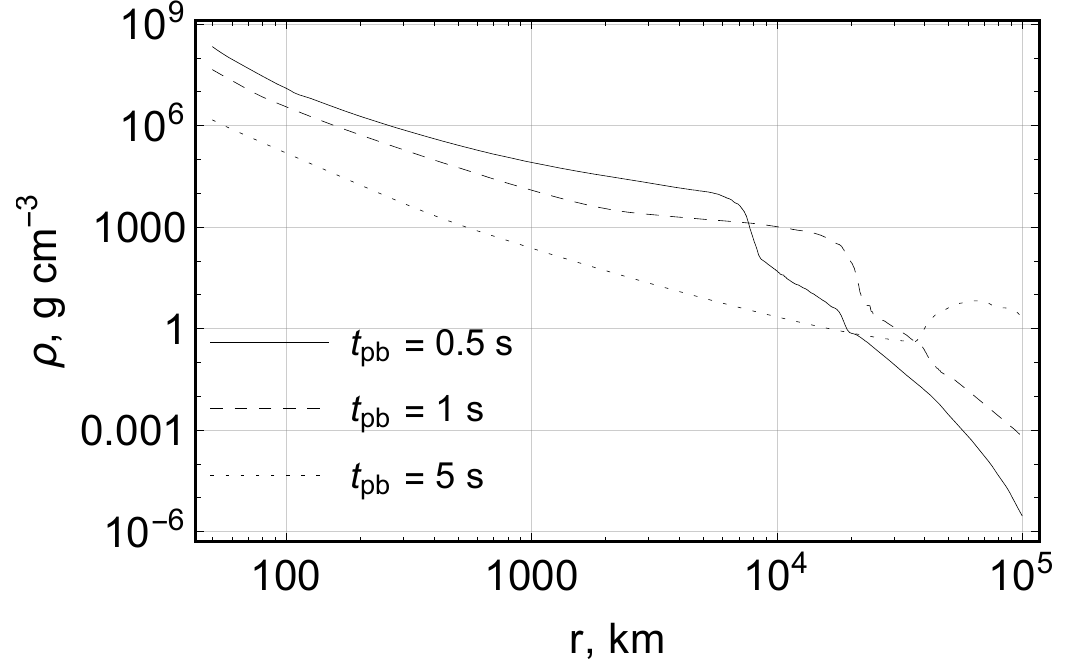}~
    \includegraphics[width=0.32 \textwidth]{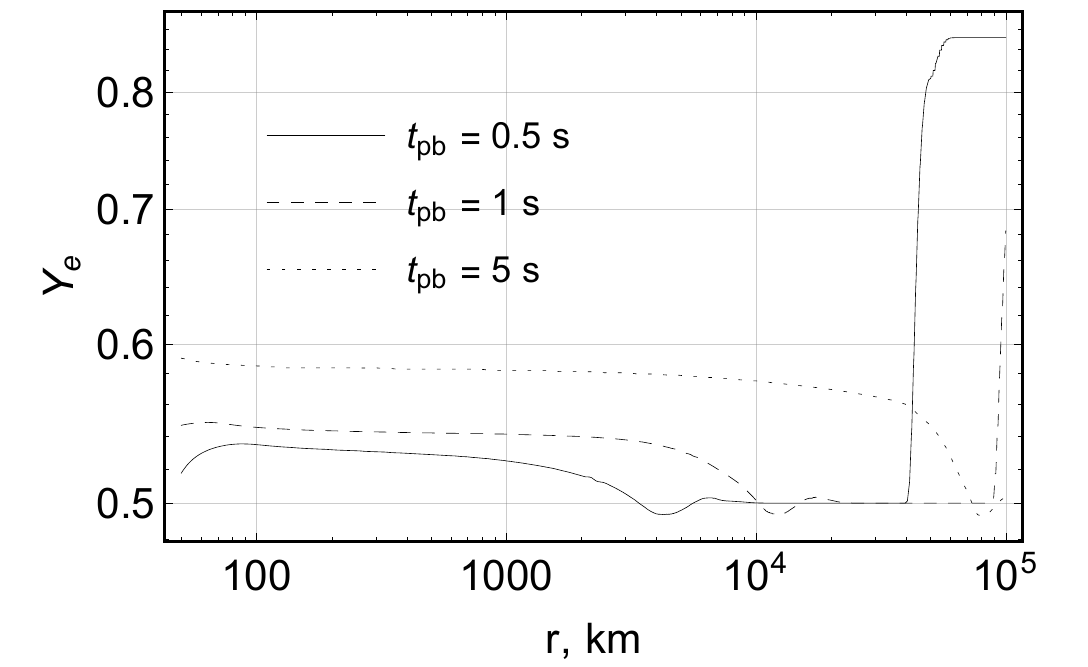}
    \caption{Temperature, density, and electron fraction $Y_e = n_e/n_B$ profiles in the reference model used to estimate the thermalization effects - \cite{2010PhRvL.104y1101H,Garchinv_archive}.
    }
    \label{fig:profiles}
\end{figure*}

To estimate the thermalization effects we used SN simulation data for a specific scenario of 8.8  $M_\odot$ simulation. The non-thermalized part of the final result remains model-dependent. Examples of the temperature, density, and electron fraction profiles are given in Fig.\ref{fig:profiles}
\label{App:Sn_profiles}
\bibliography{preamble,sample}

\let\jnlstyle=\rm\def\jref#1{{\jnlstyle#1}}\def\aj{\jref{AJ}} \def\araa{\jref{ARA\&A}} \def\apj{\jref{ApJ}\ } \def\apjl{\jref{ApJ}\ } \def\apjs{\jref{ApJS}} \def\ao{\jref{Appl.~Opt.}} \def\apss{\jref{Ap\&SS}} \def\aap{\jref{A\&A}} \def\aapr{\jref{A\&A~Rev.}} \def\aaps{\jref{A\&AS}} \def\azh{\jref{AZh}} \def\baas{\jref{BAAS}} \def\jrasc{\jref{JRASC}} \def\memras{\jref{MmRAS}} \def\mnras{\jref{MNRAS}\ } \def\pra{\jref{Phys.~Rev.~A}\ } \def\prb{\jref{Phys.~Rev.~B}\ } \def\prc{\jref{Phys.~Rev.~C}\ } \def\prd{\jref{Phys.~Rev.~D}\ } \def\pre{\jref{Phys.~Rev.~E}} \def\prl{\jref{Phys.~Rev.~Lett.}} \def\pasp{\jref{PASP}} \def\pasj{\jref{PASJ}} \def\qjras{\jref{QJRAS}} \def\skytel{\jref{S\&T}} \def\solphys{\jref{Sol.~Phys.}} \def\sovast{\jref{Soviet~Ast.}} \def\ssr{\jref{Space~Sci.~Rev.}} \def\zap{\jref{ZAp}} \def\nat{\jref{Nature}\ } \def\iaucirc{\jref{IAU~Circ.}} \def\aplett{\jref{Astrophys.~Lett.}} \def\apspr{\jref{Astrophys.~Space~Phys.~Res.}} \def\bain{\jref{Bull.~Astron.~Inst.~Netherlands}}
  \def\fcp{\jref{Fund.~Cosmic~Phys.}} \def\gca{\jref{Geochim.~Cosmochim.~Acta}} \def\grl{\jref{Geophys.~Res.~Lett.}} \def\jcp{\jref{J.~Chem.~Phys.}} \def\jgr{\jref{J.~Geophys.~Res.}} \def\jqsrt{\jref{J.~Quant.~Spec.~Radiat.~Transf.}} \def\memsai{\jref{Mem.~Soc.~Astron.~Italiana}} \def\nphysa{\jref{Nucl.~Phys.~A}} \def\physrep{\jref{Phys.~Rep.}} \def\physscr{\jref{Phys.~Scr}} \def\planss{\jref{Planet.~Space~Sci.}} \def\procspie{\jref{Proc.~SPIE}} \let\astap=\aap \let\apjlett=\apjl \let\apjsupp=\apjs \let\applopt=\ao \def\jcap{\jref{JCAP}}
\begin{thebibliography}{51}%
\makeatletter
\providecommand \@ifxundefined [1]{%
 \@ifx{#1\undefined}
}%
\providecommand \@ifnum [1]{%
 \ifnum #1\expandafter \@firstoftwo
 \else \expandafter \@secondoftwo
 \fi
}%
\providecommand \@ifx [1]{%
 \ifx #1\expandafter \@firstoftwo
 \else \expandafter \@secondoftwo
 \fi
}%
\providecommand \natexlab [1]{#1}%
\providecommand \enquote  [1]{``#1''}%
\providecommand \bibnamefont  [1]{#1}%
\providecommand \bibfnamefont [1]{#1}%
\providecommand \citenamefont [1]{#1}%
\providecommand \href@noop [0]{\@secondoftwo}%
\providecommand \href [0]{\begingroup \@sanitize@url \@href}%
\providecommand \@href[1]{\@@startlink{#1}\@@href}%
\providecommand \@@href[1]{\endgroup#1\@@endlink}%
\providecommand \@sanitize@url [0]{\catcode `\\12\catcode `\$12\catcode `\&12\catcode `\#12\catcode `\^12\catcode `\_12\catcode `\%12\relax}%
\providecommand \@@startlink[1]{}%
\providecommand \@@endlink[0]{}%
\providecommand \url  [0]{\begingroup\@sanitize@url \@url }%
\providecommand \@url [1]{\endgroup\@href {#1}{\urlprefix }}%
\providecommand \urlprefix  [0]{URL }%
\providecommand \Eprint [0]{\href }%
\providecommand \doibase [0]{http://dx.doi.org/}%
\providecommand \selectlanguage [0]{\@gobble}%
\providecommand \bibinfo  [0]{\@secondoftwo}%
\providecommand \bibfield  [0]{\@secondoftwo}%
\providecommand \translation [1]{[#1]}%
\providecommand \BibitemOpen [0]{}%
\providecommand \bibitemStop [0]{}%
\providecommand \bibitemNoStop [0]{.\EOS\space}%
\providecommand \EOS [0]{\spacefactor3000\relax}%
\providecommand \BibitemShut  [1]{\csname bibitem#1\endcsname}%
\let\auto@bib@innerbib\@empty
\bibitem [{\citenamefont {{Janka}}(2017)}]{Janka:2017vlw}%
  \BibitemOpen
  \bibfield  {author} {\bibinfo {author} {\bibfnamefont {H.-T.}\ \bibnamefont {{Janka}}},\ }\enquote {\bibinfo {title} {{Neutrino Emission from Supernovae}},}\ in\ \href {\doibase 10.1007/978-3-319-21846-5_4} {\emph {\bibinfo {booktitle} {Handbook of Supernovae}}},\ \bibinfo {editor} {edited by\ \bibinfo {editor} {\bibfnamefont {A.~W.}\ \bibnamefont {{Alsabti}}}\ and\ \bibinfo {editor} {\bibfnamefont {P.}~\bibnamefont {{Murdin}}}}\ (\bibinfo  {publisher} {{Springer International Publishing AG}},\ \bibinfo {year} {2017})\ p.\ \bibinfo {pages} {1575},\ \Eprint {http://arxiv.org/abs/1702.08713} {arXiv:1702.08713 [astro-ph.HE]} \BibitemShut {NoStop}%
\bibitem [{\citenamefont {Alsabti}\ and\ \citenamefont {Murdin}(2017)}]{Alsabti:2017ahu}%
  \BibitemOpen
  \bibinfo {editor} {\bibfnamefont {A.~W.}\ \bibnamefont {Alsabti}}\ and\ \bibinfo {editor} {\bibfnamefont {P.}~\bibnamefont {Murdin}},\ eds.,\ \href {\doibase 10.1007/978-3-319-21846-5} {\emph {\bibinfo {title} {{Handbook of Supernovae}}}}\ (\bibinfo  {publisher} {Springer},\ \bibinfo {year} {2017})\BibitemShut {NoStop}%
\bibitem [{\citenamefont {Alekhin}\ \emph {et~al.}(2016)\citenamefont {Alekhin} \emph {et~al.}}]{Alekhin:2015byh}%
  \BibitemOpen
  \bibfield  {author} {\bibinfo {author} {\bibfnamefont {S.}~\bibnamefont {Alekhin}} \emph {et~al.},\ }\href {\doibase 10.1088/0034-4885/79/12/124201} {\bibfield  {journal} {\bibinfo  {journal} {Rept. Prog. Phys.}\ }\textbf {\bibinfo {volume} {79}},\ \bibinfo {pages} {124201} (\bibinfo {year} {2016})},\ \Eprint {http://arxiv.org/abs/1504.04855} {arXiv:1504.04855 [hep-ph]} \BibitemShut {NoStop}%
\bibitem [{\citenamefont {Lanfranchi}\ \emph {et~al.}(2021)\citenamefont {Lanfranchi}, \citenamefont {Pospelov},\ and\ \citenamefont {Schuster}}]{Lanfranchi:2020crw}%
  \BibitemOpen
  \bibfield  {author} {\bibinfo {author} {\bibfnamefont {G.}~\bibnamefont {Lanfranchi}}, \bibinfo {author} {\bibfnamefont {M.}~\bibnamefont {Pospelov}}, \ and\ \bibinfo {author} {\bibfnamefont {P.}~\bibnamefont {Schuster}},\ }\href {\doibase 10.1146/annurev-nucl-102419-055056} {\bibfield  {journal} {\bibinfo  {journal} {Ann. Rev. Nucl. Part. Sci.}\ }\textbf {\bibinfo {volume} {71}},\ \bibinfo {pages} {279} (\bibinfo {year} {2021})},\ \Eprint {http://arxiv.org/abs/2011.02157} {arXiv:2011.02157 [hep-ph]} \BibitemShut {NoStop}%
\bibitem [{\citenamefont {Agrawal}\ \emph {et~al.}(2021)\citenamefont {Agrawal} \emph {et~al.}}]{Agrawal:2021dbo}%
  \BibitemOpen
  \bibfield  {author} {\bibinfo {author} {\bibfnamefont {P.}~\bibnamefont {Agrawal}} \emph {et~al.},\ }\href {\doibase 10.1140/epjc/s10052-021-09703-7} {\bibfield  {journal} {\bibinfo  {journal} {Eur. Phys. J. C}\ }\textbf {\bibinfo {volume} {81}},\ \bibinfo {pages} {1015} (\bibinfo {year} {2021})},\ \Eprint {http://arxiv.org/abs/2102.12143} {arXiv:2102.12143 [hep-ph]} \BibitemShut {NoStop}%
\bibitem [{\citenamefont {Antel}\ \emph {et~al.}(2023)\citenamefont {Antel} \emph {et~al.}}]{Antel:2023hkf}%
  \BibitemOpen
  \bibfield  {author} {\bibinfo {author} {\bibfnamefont {C.}~\bibnamefont {Antel}} \emph {et~al.},\ }in\ \href@noop {} {\emph {\bibinfo {booktitle} {{Workshop on Feebly-Interacting Particles}}}}\ (\bibinfo {year} {2023})\ \Eprint {http://arxiv.org/abs/2305.01715} {arXiv:2305.01715 [hep-ph]} \BibitemShut {NoStop}%
\bibitem [{\citenamefont {Raffelt}(1996)}]{Raffelt:1996wa}%
  \BibitemOpen
  \bibfield  {author} {\bibinfo {author} {\bibfnamefont {G.~G.}\ \bibnamefont {Raffelt}},\ }\href {http://wwwth.mpp.mpg.de/members/raffelt/mypapers/199613.pdf} {\emph {\bibinfo {title} {{Stars as laboratories for fundamental physics}}}}\ (\bibinfo  {publisher} {Chicago, USA: Univ. Pr. (1996) 664 p},\ \bibinfo {year} {1996})\BibitemShut {NoStop}%
\bibitem [{\citenamefont {Hirata}\ \emph {et~al.}(1987)\citenamefont {Hirata} \emph {et~al.}}]{Hirata:1987hu}%
  \BibitemOpen
  \bibfield  {author} {\bibinfo {author} {\bibfnamefont {K.}~\bibnamefont {Hirata}} \emph {et~al.} (\bibinfo {collaboration} {Kamiokande-II}),\ }\bibfield  {booktitle} {\emph {\bibinfo {booktitle} {{GRAND UNIFICATION. PROCEEDINGS, 8TH WORKSHOP, SYRACUSE, USA, APRIL 16-18, 1987}}},\ }\href {\doibase 10.1103/PhysRevLett.58.1490} {\bibfield  {journal} {\bibinfo  {journal} {Phys. Rev. Lett.}\ }\textbf {\bibinfo {volume} {58}},\ \bibinfo {pages} {1490} (\bibinfo {year} {1987})},\ \bibinfo {note} {[,727(1987)]}\BibitemShut {NoStop}%
\bibitem [{\citenamefont {{Svoboda}}\ \emph {et~al.}(1987)\citenamefont {{Svoboda}}, \citenamefont {{Bratton}}, \citenamefont {{Casper}}, \citenamefont {{Ciocio}},\ and\ \citenamefont {{Claus}}}]{1987ESOC...26..229S}%
  \BibitemOpen
  \bibfield  {author} {\bibinfo {author} {\bibfnamefont {R.}~\bibnamefont {{Svoboda}}}, \bibinfo {author} {\bibfnamefont {C.~B.}\ \bibnamefont {{Bratton}}}, \bibinfo {author} {\bibfnamefont {D.}~\bibnamefont {{Casper}}}, \bibinfo {author} {\bibfnamefont {A.}~\bibnamefont {{Ciocio}}}, \ and\ \bibinfo {author} {\bibfnamefont {R.}~\bibnamefont {{Claus}}},\ }in\ \href@noop {} {\emph {\bibinfo {booktitle} {European Southern Observatory Conference and Workshop Proceedings}}},\ \bibinfo {series} {European Southern Observatory Conference and Workshop Proceedings}, Vol.~\bibinfo {volume} {26},\ \bibinfo {editor} {edited by\ \bibinfo {editor} {\bibfnamefont {I.~J.}\ \bibnamefont {{Danziger}}}}\ (\bibinfo {year} {1987})\ pp.\ \bibinfo {pages} {229--235}\BibitemShut {NoStop}%
\bibitem [{\citenamefont {{Alekseev}}\ \emph {et~al.}(1987)\citenamefont {{Alekseev}}, \citenamefont {{Alekseeva}}, \citenamefont {{Krivosheina}},\ and\ \citenamefont {{Volchenko}}}]{1987ESOC...26..237A}%
  \BibitemOpen
  \bibfield  {author} {\bibinfo {author} {\bibfnamefont {E.~N.}\ \bibnamefont {{Alekseev}}}, \bibinfo {author} {\bibfnamefont {L.~N.}\ \bibnamefont {{Alekseeva}}}, \bibinfo {author} {\bibfnamefont {I.~V.}\ \bibnamefont {{Krivosheina}}}, \ and\ \bibinfo {author} {\bibfnamefont {V.~I.}\ \bibnamefont {{Volchenko}}},\ }in\ \href@noop {} {\emph {\bibinfo {booktitle} {European Southern Observatory Conference and Workshop Proceedings}}},\ \bibinfo {series} {European Southern Observatory Conference and Workshop Proceedings}, Vol.~\bibinfo {volume} {26},\ \bibinfo {editor} {edited by\ \bibinfo {editor} {\bibfnamefont {I.~J.}\ \bibnamefont {{Danziger}}}}\ (\bibinfo {year} {1987})\ pp.\ \bibinfo {pages} {237--247}\BibitemShut {NoStop}%
\bibitem [{\citenamefont {De~la Torre~Luque}\ \emph {et~al.}(2023)\citenamefont {De~la Torre~Luque}, \citenamefont {Balaji},\ and\ \citenamefont {Carenza}}]{DelaTorreLuque:2023huu}%
  \BibitemOpen
  \bibfield  {author} {\bibinfo {author} {\bibfnamefont {P.}~\bibnamefont {De~la Torre~Luque}}, \bibinfo {author} {\bibfnamefont {S.}~\bibnamefont {Balaji}}, \ and\ \bibinfo {author} {\bibfnamefont {P.}~\bibnamefont {Carenza}},\ }\href@noop {} {\  (\bibinfo {year} {2023})},\ \Eprint {http://arxiv.org/abs/2307.13731} {arXiv:2307.13731 [hep-ph]} \BibitemShut {NoStop}%
\bibitem [{\citenamefont {Hoof}\ and\ \citenamefont {Schulz}(2023)}]{Hoof:2022xbe}%
  \BibitemOpen
  \bibfield  {author} {\bibinfo {author} {\bibfnamefont {S.}~\bibnamefont {Hoof}}\ and\ \bibinfo {author} {\bibfnamefont {L.}~\bibnamefont {Schulz}},\ }\href {\doibase 10.1088/1475-7516/2023/03/054} {\bibfield  {journal} {\bibinfo  {journal} {JCAP}\ }\textbf {\bibinfo {volume} {03}},\ \bibinfo {pages} {054} (\bibinfo {year} {2023})},\ \Eprint {http://arxiv.org/abs/2212.09764} {arXiv:2212.09764 [hep-ph]} \BibitemShut {NoStop}%
\bibitem [{\citenamefont {Dev}\ \emph {et~al.}(2020)\citenamefont {Dev}, \citenamefont {Mohapatra},\ and\ \citenamefont {Zhang}}]{Dev:2020eam}%
  \BibitemOpen
  \bibfield  {author} {\bibinfo {author} {\bibfnamefont {P.~S.~B.}\ \bibnamefont {Dev}}, \bibinfo {author} {\bibfnamefont {R.~N.}\ \bibnamefont {Mohapatra}}, \ and\ \bibinfo {author} {\bibfnamefont {Y.}~\bibnamefont {Zhang}},\ }\href {\doibase 10.1088/1475-7516/2020/08/003} {\bibfield  {journal} {\bibinfo  {journal} {JCAP}\ }\textbf {\bibinfo {volume} {08}},\ \bibinfo {pages} {003} (\bibinfo {year} {2020})},\ \bibinfo {note} {[Erratum: JCAP 11, E01 (2020)]},\ \Eprint {http://arxiv.org/abs/2005.00490} {arXiv:2005.00490 [hep-ph]} \BibitemShut {NoStop}%
\bibitem [{\citenamefont {Chen}\ \emph {et~al.}(2022)\citenamefont {Chen}, \citenamefont {Sen}, \citenamefont {Tangarife}, \citenamefont {Tuckler},\ and\ \citenamefont {Zhang}}]{Chen:2022kal}%
  \BibitemOpen
  \bibfield  {author} {\bibinfo {author} {\bibfnamefont {Y.-M.}\ \bibnamefont {Chen}}, \bibinfo {author} {\bibfnamefont {M.}~\bibnamefont {Sen}}, \bibinfo {author} {\bibfnamefont {W.}~\bibnamefont {Tangarife}}, \bibinfo {author} {\bibfnamefont {D.}~\bibnamefont {Tuckler}}, \ and\ \bibinfo {author} {\bibfnamefont {Y.}~\bibnamefont {Zhang}},\ }\href {\doibase 10.1088/1475-7516/2022/11/014} {\bibfield  {journal} {\bibinfo  {journal} {JCAP}\ }\textbf {\bibinfo {volume} {11}},\ \bibinfo {pages} {014} (\bibinfo {year} {2022})},\ \Eprint {http://arxiv.org/abs/2207.14300} {arXiv:2207.14300 [hep-ph]} \BibitemShut {NoStop}%
\bibitem [{\citenamefont {Magill}\ \emph {et~al.}(2018)\citenamefont {Magill}, \citenamefont {Plestid}, \citenamefont {Pospelov},\ and\ \citenamefont {Tsai}}]{Magill:2018jla}%
  \BibitemOpen
  \bibfield  {author} {\bibinfo {author} {\bibfnamefont {G.}~\bibnamefont {Magill}}, \bibinfo {author} {\bibfnamefont {R.}~\bibnamefont {Plestid}}, \bibinfo {author} {\bibfnamefont {M.}~\bibnamefont {Pospelov}}, \ and\ \bibinfo {author} {\bibfnamefont {Y.-D.}\ \bibnamefont {Tsai}},\ }\href {\doibase 10.1103/PhysRevD.98.115015} {\bibfield  {journal} {\bibinfo  {journal} {Phys. Rev. D}\ }\textbf {\bibinfo {volume} {98}},\ \bibinfo {pages} {115015} (\bibinfo {year} {2018})},\ \Eprint {http://arxiv.org/abs/1803.03262} {arXiv:1803.03262 [hep-ph]} \BibitemShut {NoStop}%
\bibitem [{\citenamefont {Heurtier}\ and\ \citenamefont {Zhang}(2017)}]{Heurtier:2016otg}%
  \BibitemOpen
  \bibfield  {author} {\bibinfo {author} {\bibfnamefont {L.}~\bibnamefont {Heurtier}}\ and\ \bibinfo {author} {\bibfnamefont {Y.}~\bibnamefont {Zhang}},\ }\href {\doibase 10.1088/1475-7516/2017/02/042} {\bibfield  {journal} {\bibinfo  {journal} {JCAP}\ }\textbf {\bibinfo {volume} {1702}},\ \bibinfo {pages} {042} (\bibinfo {year} {2017})},\ \Eprint {http://arxiv.org/abs/1609.05882} {arXiv:1609.05882 [hep-ph]} \BibitemShut {NoStop}%
\bibitem [{\citenamefont {Kazanas}\ \emph {et~al.}(2014)\citenamefont {Kazanas}, \citenamefont {Mohapatra}, \citenamefont {Nussinov}, \citenamefont {Teplitz},\ and\ \citenamefont {Zhang}}]{Kazanas:2014mca}%
  \BibitemOpen
  \bibfield  {author} {\bibinfo {author} {\bibfnamefont {D.}~\bibnamefont {Kazanas}}, \bibinfo {author} {\bibfnamefont {R.~N.}\ \bibnamefont {Mohapatra}}, \bibinfo {author} {\bibfnamefont {S.}~\bibnamefont {Nussinov}}, \bibinfo {author} {\bibfnamefont {V.~L.}\ \bibnamefont {Teplitz}}, \ and\ \bibinfo {author} {\bibfnamefont {Y.}~\bibnamefont {Zhang}},\ }\href {\doibase 10.1016/j.nuclphysb.2014.11.009} {\bibfield  {journal} {\bibinfo  {journal} {Nucl. Phys. B}\ }\textbf {\bibinfo {volume} {890}},\ \bibinfo {pages} {17} (\bibinfo {year} {2014})},\ \Eprint {http://arxiv.org/abs/1410.0221} {arXiv:1410.0221 [hep-ph]} \BibitemShut {NoStop}%
\bibitem [{\citenamefont {Chang}\ \emph {et~al.}(2017)\citenamefont {Chang}, \citenamefont {Essig},\ and\ \citenamefont {McDermott}}]{Chang:2016ntp}%
  \BibitemOpen
  \bibfield  {author} {\bibinfo {author} {\bibfnamefont {J.~H.}\ \bibnamefont {Chang}}, \bibinfo {author} {\bibfnamefont {R.}~\bibnamefont {Essig}}, \ and\ \bibinfo {author} {\bibfnamefont {S.~D.}\ \bibnamefont {McDermott}},\ }\href {\doibase 10.1007/JHEP01(2017)107} {\bibfield  {journal} {\bibinfo  {journal} {JHEP}\ }\textbf {\bibinfo {volume} {01}},\ \bibinfo {pages} {107} (\bibinfo {year} {2017})},\ \Eprint {http://arxiv.org/abs/1611.03864} {arXiv:1611.03864 [hep-ph]} \BibitemShut {NoStop}%
\bibitem [{\citenamefont {Hook}\ \emph {et~al.}(2021)\citenamefont {Hook}, \citenamefont {Marques-Tavares},\ and\ \citenamefont {Ristow}}]{Hook:2021ous}%
  \BibitemOpen
  \bibfield  {author} {\bibinfo {author} {\bibfnamefont {A.}~\bibnamefont {Hook}}, \bibinfo {author} {\bibfnamefont {G.}~\bibnamefont {Marques-Tavares}}, \ and\ \bibinfo {author} {\bibfnamefont {C.}~\bibnamefont {Ristow}},\ }\href {\doibase 10.1007/JHEP06(2021)167} {\bibfield  {journal} {\bibinfo  {journal} {JHEP}\ }\textbf {\bibinfo {volume} {06}},\ \bibinfo {pages} {167} (\bibinfo {year} {2021})},\ \Eprint {http://arxiv.org/abs/2105.06476} {arXiv:2105.06476 [hep-ph]} \BibitemShut {NoStop}%
\bibitem [{\citenamefont {Dent}\ \emph {et~al.}(2012)\citenamefont {Dent}, \citenamefont {Ferrer},\ and\ \citenamefont {Krauss}}]{Dent:2012mx}%
  \BibitemOpen
  \bibfield  {author} {\bibinfo {author} {\bibfnamefont {J.~B.}\ \bibnamefont {Dent}}, \bibinfo {author} {\bibfnamefont {F.}~\bibnamefont {Ferrer}}, \ and\ \bibinfo {author} {\bibfnamefont {L.~M.}\ \bibnamefont {Krauss}},\ }\href@noop {} {\  (\bibinfo {year} {2012})},\ \Eprint {http://arxiv.org/abs/1201.2683} {arXiv:1201.2683 [astro-ph.CO]} \BibitemShut {NoStop}%
\bibitem [{\citenamefont {Burrows}\ \emph {et~al.}(1989)\citenamefont {Burrows}, \citenamefont {Turner},\ and\ \citenamefont {Brinkmann}}]{Burrows:1988ah}%
  \BibitemOpen
  \bibfield  {author} {\bibinfo {author} {\bibfnamefont {A.}~\bibnamefont {Burrows}}, \bibinfo {author} {\bibfnamefont {M.~S.}\ \bibnamefont {Turner}}, \ and\ \bibinfo {author} {\bibfnamefont {R.~P.}\ \bibnamefont {Brinkmann}},\ }\href {\doibase 10.1103/PhysRevD.39.1020} {\bibfield  {journal} {\bibinfo  {journal} {Phys. Rev.}\ }\textbf {\bibinfo {volume} {D39}},\ \bibinfo {pages} {1020} (\bibinfo {year} {1989})}\BibitemShut {NoStop}%
\bibitem [{\citenamefont {Burrows}\ and\ \citenamefont {Vartanyan}(2021)}]{Burrows:2020qrp}%
  \BibitemOpen
  \bibfield  {author} {\bibinfo {author} {\bibfnamefont {A.}~\bibnamefont {Burrows}}\ and\ \bibinfo {author} {\bibfnamefont {D.}~\bibnamefont {Vartanyan}},\ }\href {\doibase 10.1038/s41586-020-03059-w} {\bibfield  {journal} {\bibinfo  {journal} {Nature}\ }\textbf {\bibinfo {volume} {589}},\ \bibinfo {pages} {29} (\bibinfo {year} {2021})},\ \Eprint {http://arxiv.org/abs/2009.14157} {arXiv:2009.14157 [astro-ph.SR]} \BibitemShut {NoStop}%
\bibitem [{\citenamefont {Fiorillo}\ \emph {et~al.}(2023)\citenamefont {Fiorillo}, \citenamefont {Raffelt},\ and\ \citenamefont {Vitagliano}}]{Fiorillo:2022cdq}%
  \BibitemOpen
  \bibfield  {author} {\bibinfo {author} {\bibfnamefont {D.~F.~G.}\ \bibnamefont {Fiorillo}}, \bibinfo {author} {\bibfnamefont {G.~G.}\ \bibnamefont {Raffelt}}, \ and\ \bibinfo {author} {\bibfnamefont {E.}~\bibnamefont {Vitagliano}},\ }\href {\doibase 10.1103/PhysRevLett.131.021001} {\bibfield  {journal} {\bibinfo  {journal} {Phys. Rev. Lett.}\ }\textbf {\bibinfo {volume} {131}},\ \bibinfo {pages} {021001} (\bibinfo {year} {2023})},\ \Eprint {http://arxiv.org/abs/2209.11773} {arXiv:2209.11773 [hep-ph]} \BibitemShut {NoStop}%
\bibitem [{\citenamefont {Akita}\ \emph {et~al.}(2022)\citenamefont {Akita}, \citenamefont {Im},\ and\ \citenamefont {Masud}}]{Akita:2022etk}%
  \BibitemOpen
  \bibfield  {author} {\bibinfo {author} {\bibfnamefont {K.}~\bibnamefont {Akita}}, \bibinfo {author} {\bibfnamefont {S.~H.}\ \bibnamefont {Im}}, \ and\ \bibinfo {author} {\bibfnamefont {M.}~\bibnamefont {Masud}},\ }\href {\doibase 10.1007/JHEP12(2022)050} {\bibfield  {journal} {\bibinfo  {journal} {JHEP}\ }\textbf {\bibinfo {volume} {12}},\ \bibinfo {pages} {050} (\bibinfo {year} {2022})},\ \Eprint {http://arxiv.org/abs/2206.06852} {arXiv:2206.06852 [hep-ph]} \BibitemShut {NoStop}%
\bibitem [{\citenamefont {Syvolap}(2023)}]{Syvolap:2023trc}%
  \BibitemOpen
  \bibfield  {author} {\bibinfo {author} {\bibfnamefont {V.}~\bibnamefont {Syvolap}},\ }\href@noop {} {\  (\bibinfo {year} {2023})},\ \Eprint {http://arxiv.org/abs/2301.07052} {arXiv:2301.07052 [hep-ph]} \BibitemShut {NoStop}%
\bibitem [{\citenamefont {Okun}(1982)}]{Okun:1982xi}%
  \BibitemOpen
  \bibfield  {author} {\bibinfo {author} {\bibfnamefont {L.~B.}\ \bibnamefont {Okun}},\ }\href@noop {} {\bibfield  {journal} {\bibinfo  {journal} {Sov. Phys. JETP}\ }\textbf {\bibinfo {volume} {56}},\ \bibinfo {pages} {502} (\bibinfo {year} {1982})}\BibitemShut {NoStop}%
\bibitem [{\citenamefont {Holdom}(1986)}]{Holdom:1985ag}%
  \BibitemOpen
  \bibfield  {author} {\bibinfo {author} {\bibfnamefont {B.}~\bibnamefont {Holdom}},\ }\href {\doibase 10.1016/0370-2693(86)91377-8} {\bibfield  {journal} {\bibinfo  {journal} {Phys. Lett. B}\ }\textbf {\bibinfo {volume} {166}},\ \bibinfo {pages} {196} (\bibinfo {year} {1986})}\BibitemShut {NoStop}%
\bibitem [{\citenamefont {Fabbrichesi}\ \emph {et~al.}(2020)\citenamefont {Fabbrichesi}, \citenamefont {Gabrielli},\ and\ \citenamefont {Lanfranchi}}]{Fabbrichesi:2020wbt}%
  \BibitemOpen
  \bibfield  {author} {\bibinfo {author} {\bibfnamefont {M.}~\bibnamefont {Fabbrichesi}}, \bibinfo {author} {\bibfnamefont {E.}~\bibnamefont {Gabrielli}}, \ and\ \bibinfo {author} {\bibfnamefont {G.}~\bibnamefont {Lanfranchi}},\ }\href {\doibase 10.1007/978-3-030-62519-1} {\  (\bibinfo {year} {2020}),\ 10.1007/978-3-030-62519-1},\ \Eprint {http://arxiv.org/abs/2005.01515} {arXiv:2005.01515 [hep-ph]} \BibitemShut {NoStop}%
\bibitem [{\citenamefont {Caputo}\ \emph {et~al.}(2021)\citenamefont {Caputo}, \citenamefont {Millar}, \citenamefont {O'Hare},\ and\ \citenamefont {Vitagliano}}]{Caputo:2021eaa}%
  \BibitemOpen
  \bibfield  {author} {\bibinfo {author} {\bibfnamefont {A.}~\bibnamefont {Caputo}}, \bibinfo {author} {\bibfnamefont {A.~J.}\ \bibnamefont {Millar}}, \bibinfo {author} {\bibfnamefont {C.~A.~J.}\ \bibnamefont {O'Hare}}, \ and\ \bibinfo {author} {\bibfnamefont {E.}~\bibnamefont {Vitagliano}},\ }\href {\doibase 10.1103/PhysRevD.104.095029} {\bibfield  {journal} {\bibinfo  {journal} {Phys. Rev. D}\ }\textbf {\bibinfo {volume} {104}},\ \bibinfo {pages} {095029} (\bibinfo {year} {2021})},\ \Eprint {http://arxiv.org/abs/2105.04565} {arXiv:2105.04565 [hep-ph]} \BibitemShut {NoStop}%
\bibitem [{\citenamefont {DeRocco}\ \emph {et~al.}(2019)\citenamefont {DeRocco}, \citenamefont {Graham}, \citenamefont {Kasen}, \citenamefont {Marques-Tavares},\ and\ \citenamefont {Rajendran}}]{DeRocco:2019njg}%
  \BibitemOpen
  \bibfield  {author} {\bibinfo {author} {\bibfnamefont {W.}~\bibnamefont {DeRocco}}, \bibinfo {author} {\bibfnamefont {P.~W.}\ \bibnamefont {Graham}}, \bibinfo {author} {\bibfnamefont {D.}~\bibnamefont {Kasen}}, \bibinfo {author} {\bibfnamefont {G.}~\bibnamefont {Marques-Tavares}}, \ and\ \bibinfo {author} {\bibfnamefont {S.}~\bibnamefont {Rajendran}},\ }\href {\doibase 10.1007/JHEP02(2019)171} {\bibfield  {journal} {\bibinfo  {journal} {JHEP}\ }\textbf {\bibinfo {volume} {02}},\ \bibinfo {pages} {171} (\bibinfo {year} {2019})},\ \Eprint {http://arxiv.org/abs/1901.08596} {arXiv:1901.08596 [hep-ph]} \BibitemShut {NoStop}%
\bibitem [{\citenamefont {Mastrototaro}\ \emph {et~al.}(2020)\citenamefont {Mastrototaro}, \citenamefont {Mirizzi}, \citenamefont {Serpico},\ and\ \citenamefont {Esmaili}}]{Mastrototaro:2019vug}%
  \BibitemOpen
  \bibfield  {author} {\bibinfo {author} {\bibfnamefont {L.}~\bibnamefont {Mastrototaro}}, \bibinfo {author} {\bibfnamefont {A.}~\bibnamefont {Mirizzi}}, \bibinfo {author} {\bibfnamefont {P.~D.}\ \bibnamefont {Serpico}}, \ and\ \bibinfo {author} {\bibfnamefont {A.}~\bibnamefont {Esmaili}},\ }\href {\doibase 10.1088/1475-7516/2020/01/010} {\bibfield  {journal} {\bibinfo  {journal} {JCAP}\ }\textbf {\bibinfo {volume} {01}},\ \bibinfo {pages} {010} (\bibinfo {year} {2020})},\ \Eprint {http://arxiv.org/abs/1910.10249} {arXiv:1910.10249 [hep-ph]} \BibitemShut {NoStop}%
\bibitem [{\citenamefont {Abe}\ \emph {et~al.}(2018)\citenamefont {Abe} \emph {et~al.}}]{Hyper-Kamiokande:2016srs}%
  \BibitemOpen
  \bibfield  {author} {\bibinfo {author} {\bibfnamefont {K.}~\bibnamefont {Abe}} \emph {et~al.} (\bibinfo {collaboration} {Hyper-Kamiokande}),\ }\href {\doibase 10.1093/ptep/pty044} {\bibfield  {journal} {\bibinfo  {journal} {PTEP}\ }\textbf {\bibinfo {volume} {2018}},\ \bibinfo {pages} {063C01} (\bibinfo {year} {2018})},\ \Eprint {http://arxiv.org/abs/1611.06118} {arXiv:1611.06118 [hep-ex]} \BibitemShut {NoStop}%
\bibitem [{\citenamefont {Abusleme}\ \emph {et~al.}(2022)\citenamefont {Abusleme} \emph {et~al.}}]{JUNO:2021vlw}%
  \BibitemOpen
  \bibfield  {author} {\bibinfo {author} {\bibfnamefont {A.}~\bibnamefont {Abusleme}} \emph {et~al.} (\bibinfo {collaboration} {JUNO}),\ }\href {\doibase 10.1016/j.ppnp.2021.103927} {\bibfield  {journal} {\bibinfo  {journal} {Prog. Part. Nucl. Phys.}\ }\textbf {\bibinfo {volume} {123}},\ \bibinfo {pages} {103927} (\bibinfo {year} {2022})},\ \Eprint {http://arxiv.org/abs/2104.02565} {arXiv:2104.02565 [hep-ex]} \BibitemShut {NoStop}%
\bibitem [{\citenamefont {Hewes}\ \emph {et~al.}(2021)\citenamefont {Hewes} \emph {et~al.}}]{DUNE:2021tad}%
  \BibitemOpen
  \bibfield  {author} {\bibinfo {author} {\bibfnamefont {V.}~\bibnamefont {Hewes}} \emph {et~al.} (\bibinfo {collaboration} {DUNE}),\ }\href {\doibase 10.3390/instruments5040031} {\bibfield  {journal} {\bibinfo  {journal} {Instruments}\ }\textbf {\bibinfo {volume} {5}},\ \bibinfo {pages} {31} (\bibinfo {year} {2021})},\ \Eprint {http://arxiv.org/abs/2103.13910} {arXiv:2103.13910 [physics.ins-det]} \BibitemShut {NoStop}%
\bibitem [{\citenamefont {Ericson}\ \emph {et~al.}(2002)\citenamefont {Ericson}, \citenamefont {Loiseau},\ and\ \citenamefont {Thomas}}]{Ericson:2000md}%
  \BibitemOpen
  \bibfield  {author} {\bibinfo {author} {\bibfnamefont {T.~E.~O.}\ \bibnamefont {Ericson}}, \bibinfo {author} {\bibfnamefont {B.}~\bibnamefont {Loiseau}}, \ and\ \bibinfo {author} {\bibfnamefont {A.~W.}\ \bibnamefont {Thomas}},\ }\href {\doibase 10.1103/PhysRevC.66.014005} {\bibfield  {journal} {\bibinfo  {journal} {Phys. Rev. C}\ }\textbf {\bibinfo {volume} {66}},\ \bibinfo {pages} {014005} (\bibinfo {year} {2002})},\ \Eprint {http://arxiv.org/abs/hep-ph/0009312} {arXiv:hep-ph/0009312} \BibitemShut {NoStop}%
\bibitem [{\citenamefont {Oberauer}\ \emph {et~al.}(1993)\citenamefont {Oberauer}, \citenamefont {Hagner}, \citenamefont {Raffelt},\ and\ \citenamefont {Rieger}}]{Oberauer:1993yr}%
  \BibitemOpen
  \bibfield  {author} {\bibinfo {author} {\bibfnamefont {L.}~\bibnamefont {Oberauer}}, \bibinfo {author} {\bibfnamefont {C.}~\bibnamefont {Hagner}}, \bibinfo {author} {\bibfnamefont {G.}~\bibnamefont {Raffelt}}, \ and\ \bibinfo {author} {\bibfnamefont {E.}~\bibnamefont {Rieger}},\ }\href {\doibase 10.1016/0927-6505(93)90004-W} {\bibfield  {journal} {\bibinfo  {journal} {Astropart. Phys.}\ }\textbf {\bibinfo {volume} {1}},\ \bibinfo {pages} {377} (\bibinfo {year} {1993})}\BibitemShut {NoStop}%
\bibitem [{\citenamefont {Groom}\ \emph {et~al.}(2001)\citenamefont {Groom}, \citenamefont {Mokhov},\ and\ \citenamefont {Striganov}}]{Groom:2001kq}%
  \BibitemOpen
  \bibfield  {author} {\bibinfo {author} {\bibfnamefont {D.~E.}\ \bibnamefont {Groom}}, \bibinfo {author} {\bibfnamefont {N.~V.}\ \bibnamefont {Mokhov}}, \ and\ \bibinfo {author} {\bibfnamefont {S.~I.}\ \bibnamefont {Striganov}},\ }\href {\doibase 10.1006/adnd.2001.0861} {\bibfield  {journal} {\bibinfo  {journal} {Atom. Data Nucl. Data Tabl.}\ }\textbf {\bibinfo {volume} {78}},\ \bibinfo {pages} {183} (\bibinfo {year} {2001})}\BibitemShut {NoStop}%
\bibitem [{\citenamefont {{H{\"u}depohl}}\ \emph {et~al.}(2010)\citenamefont {{H{\"u}depohl}}, \citenamefont {{M{\"u}ller}}, \citenamefont {{Janka}}, \citenamefont {{Marek}},\ and\ \citenamefont {{Raffelt}}}]{2010PhRvL.104y1101H}%
  \BibitemOpen
  \bibfield  {author} {\bibinfo {author} {\bibfnamefont {L.}~\bibnamefont {{H{\"u}depohl}}}, \bibinfo {author} {\bibfnamefont {B.}~\bibnamefont {{M{\"u}ller}}}, \bibinfo {author} {\bibfnamefont {H.~T.}\ \bibnamefont {{Janka}}}, \bibinfo {author} {\bibfnamefont {A.}~\bibnamefont {{Marek}}}, \ and\ \bibinfo {author} {\bibfnamefont {G.~G.}\ \bibnamefont {{Raffelt}}},\ }\href {\doibase 10.1103/PhysRevLett.104.251101} {\bibfield  {journal} {\bibinfo  {journal} {\prl}\ }\textbf {\bibinfo {volume} {104}},\ \bibinfo {eid} {251101} (\bibinfo {year} {2010})},\ \Eprint {http://arxiv.org/abs/0912.0260} {arXiv:0912.0260 [astro-ph.SR]} \BibitemShut {NoStop}%
\bibitem [{Gar()}]{Garchinv_archive}%
  \BibitemOpen
  \href@noop {} {\enquote {\bibinfo {title} {{The Garching Core-Collapse Supernova Archive}},}\ }\bibinfo {note} {\url{https://wwwmpa.mpa-garching.mpg.de/ccsnarchive}}\BibitemShut {NoStop}%
\bibitem [{\citenamefont {Zyla}\ \emph {et~al.}(2020)\citenamefont {Zyla} \emph {et~al.}}]{ParticleDataGroup:2020ssz}%
  \BibitemOpen
  \bibfield  {author} {\bibinfo {author} {\bibfnamefont {P.~A.}\ \bibnamefont {Zyla}} \emph {et~al.} (\bibinfo {collaboration} {Particle Data Group}),\ }\href {\doibase 10.1093/ptep/ptaa104} {\bibfield  {journal} {\bibinfo  {journal} {PTEP}\ }\textbf {\bibinfo {volume} {2020}},\ \bibinfo {pages} {083C01} (\bibinfo {year} {2020})}\BibitemShut {NoStop}%
\bibitem [{\citenamefont {Giganti}\ \emph {et~al.}(2018)\citenamefont {Giganti}, \citenamefont {Lavignac},\ and\ \citenamefont {Zito}}]{Giganti:2017fhf}%
  \BibitemOpen
  \bibfield  {author} {\bibinfo {author} {\bibfnamefont {C.}~\bibnamefont {Giganti}}, \bibinfo {author} {\bibfnamefont {S.}~\bibnamefont {Lavignac}}, \ and\ \bibinfo {author} {\bibfnamefont {M.}~\bibnamefont {Zito}},\ }\href {\doibase 10.1016/j.ppnp.2017.10.001} {\bibfield  {journal} {\bibinfo  {journal} {Prog. Part. Nucl. Phys.}\ }\textbf {\bibinfo {volume} {98}},\ \bibinfo {pages} {1} (\bibinfo {year} {2018})},\ \Eprint {http://arxiv.org/abs/1710.00715} {arXiv:1710.00715 [hep-ex]} \BibitemShut {NoStop}%
\bibitem [{\citenamefont {Mikheev}\ and\ \citenamefont {Smirnov}(1985)}]{Mikheev:1986gs}%
  \BibitemOpen
  \bibfield  {author} {\bibinfo {author} {\bibfnamefont {S.~P.}\ \bibnamefont {Mikheev}}\ and\ \bibinfo {author} {\bibfnamefont {A.~{\relax Yu}.}\ \bibnamefont {Smirnov}},\ }\href@noop {} {\bibfield  {journal} {\bibinfo  {journal} {Sov. J. Nucl. Phys.}\ }\textbf {\bibinfo {volume} {42}},\ \bibinfo {pages} {913} (\bibinfo {year} {1985})},\ \bibinfo {note} {[Yad. Fiz.42,1441(1985)]}\BibitemShut {NoStop}%
\bibitem [{\citenamefont {Wolfenstein}(1978)}]{Wolfenstein:1977ue}%
  \BibitemOpen
  \bibfield  {author} {\bibinfo {author} {\bibfnamefont {L.}~\bibnamefont {Wolfenstein}},\ }\href {\doibase 10.1103/PhysRevD.17.2369} {\bibfield  {journal} {\bibinfo  {journal} {Phys. Rev.}\ }\textbf {\bibinfo {volume} {D17}},\ \bibinfo {pages} {2369} (\bibinfo {year} {1978})}\BibitemShut {NoStop}%
\bibitem [{\citenamefont {Mirizzi}\ \emph {et~al.}(2016)\citenamefont {Mirizzi}, \citenamefont {Tamborra}, \citenamefont {Janka}, \citenamefont {Saviano}, \citenamefont {Scholberg}, \citenamefont {Bollig}, \citenamefont {Hudepohl},\ and\ \citenamefont {Chakraborty}}]{Mirizzi:2015eza}%
  \BibitemOpen
  \bibfield  {author} {\bibinfo {author} {\bibfnamefont {A.}~\bibnamefont {Mirizzi}}, \bibinfo {author} {\bibfnamefont {I.}~\bibnamefont {Tamborra}}, \bibinfo {author} {\bibfnamefont {H.-T.}\ \bibnamefont {Janka}}, \bibinfo {author} {\bibfnamefont {N.}~\bibnamefont {Saviano}}, \bibinfo {author} {\bibfnamefont {K.}~\bibnamefont {Scholberg}}, \bibinfo {author} {\bibfnamefont {R.}~\bibnamefont {Bollig}}, \bibinfo {author} {\bibfnamefont {L.}~\bibnamefont {Hudepohl}}, \ and\ \bibinfo {author} {\bibfnamefont {S.}~\bibnamefont {Chakraborty}},\ }\href {\doibase 10.1393/ncr/i2016-10120-8} {\bibfield  {journal} {\bibinfo  {journal} {Riv. Nuovo Cim.}\ }\textbf {\bibinfo {volume} {39}},\ \bibinfo {pages} {1} (\bibinfo {year} {2016})},\ \Eprint {http://arxiv.org/abs/1508.00785} {arXiv:1508.00785 [astro-ph.HE]} \BibitemShut {NoStop}%
\bibitem [{\citenamefont {Capozzi}\ \emph {et~al.}(2014)\citenamefont {Capozzi}, \citenamefont {Fogli}, \citenamefont {Lisi}, \citenamefont {Marrone}, \citenamefont {Montanino},\ and\ \citenamefont {Palazzo}}]{Capozzi:2013csa}%
  \BibitemOpen
  \bibfield  {author} {\bibinfo {author} {\bibfnamefont {F.}~\bibnamefont {Capozzi}}, \bibinfo {author} {\bibfnamefont {G.~L.}\ \bibnamefont {Fogli}}, \bibinfo {author} {\bibfnamefont {E.}~\bibnamefont {Lisi}}, \bibinfo {author} {\bibfnamefont {A.}~\bibnamefont {Marrone}}, \bibinfo {author} {\bibfnamefont {D.}~\bibnamefont {Montanino}}, \ and\ \bibinfo {author} {\bibfnamefont {A.}~\bibnamefont {Palazzo}},\ }\href {\doibase 10.1103/PhysRevD.89.093018} {\bibfield  {journal} {\bibinfo  {journal} {Phys. Rev. D}\ }\textbf {\bibinfo {volume} {89}},\ \bibinfo {pages} {093018} (\bibinfo {year} {2014})},\ \Eprint {http://arxiv.org/abs/1312.2878} {arXiv:1312.2878 [hep-ph]} \BibitemShut {NoStop}%
\bibitem [{\citenamefont {Tamborra}\ \emph {et~al.}(2012)\citenamefont {Tamborra}, \citenamefont {Muller}, \citenamefont {Hudepohl}, \citenamefont {Janka},\ and\ \citenamefont {Raffelt}}]{Tamborra:2012ac}%
  \BibitemOpen
  \bibfield  {author} {\bibinfo {author} {\bibfnamefont {I.}~\bibnamefont {Tamborra}}, \bibinfo {author} {\bibfnamefont {B.}~\bibnamefont {Muller}}, \bibinfo {author} {\bibfnamefont {L.}~\bibnamefont {Hudepohl}}, \bibinfo {author} {\bibfnamefont {H.-T.}\ \bibnamefont {Janka}}, \ and\ \bibinfo {author} {\bibfnamefont {G.}~\bibnamefont {Raffelt}},\ }\href {\doibase 10.1103/PhysRevD.86.125031} {\bibfield  {journal} {\bibinfo  {journal} {Phys. Rev. D}\ }\textbf {\bibinfo {volume} {86}},\ \bibinfo {pages} {125031} (\bibinfo {year} {2012})},\ \Eprint {http://arxiv.org/abs/1211.3920} {arXiv:1211.3920 [astro-ph.SR]} \BibitemShut {NoStop}%
\bibitem [{\citenamefont {Batell}\ \emph {et~al.}(2014)\citenamefont {Batell}, \citenamefont {Essig},\ and\ \citenamefont {Surujon}}]{Batell:2014mga}%
  \BibitemOpen
  \bibfield  {author} {\bibinfo {author} {\bibfnamefont {B.}~\bibnamefont {Batell}}, \bibinfo {author} {\bibfnamefont {R.}~\bibnamefont {Essig}}, \ and\ \bibinfo {author} {\bibfnamefont {Z.}~\bibnamefont {Surujon}},\ }\href {\doibase 10.1103/PhysRevLett.113.171802} {\bibfield  {journal} {\bibinfo  {journal} {Phys. Rev. Lett.}\ }\textbf {\bibinfo {volume} {113}},\ \bibinfo {pages} {171802} (\bibinfo {year} {2014})},\ \Eprint {http://arxiv.org/abs/1406.2698} {arXiv:1406.2698 [hep-ph]} \BibitemShut {NoStop}%
\bibitem [{\citenamefont {Ahdida}\ \emph {et~al.}(2019)\citenamefont {Ahdida} \emph {et~al.}}]{SHiP:2018xqw}%
  \BibitemOpen
  \bibfield  {author} {\bibinfo {author} {\bibfnamefont {C.}~\bibnamefont {Ahdida}} \emph {et~al.} (\bibinfo {collaboration} {SHiP}),\ }\href {\doibase 10.1007/JHEP04(2019)077} {\bibfield  {journal} {\bibinfo  {journal} {JHEP}\ }\textbf {\bibinfo {volume} {04}},\ \bibinfo {pages} {077} (\bibinfo {year} {2019})},\ \Eprint {http://arxiv.org/abs/1811.00930} {arXiv:1811.00930 [hep-ph]} \BibitemShut {NoStop}%
\bibitem [{\citenamefont {Watanabe}\ \emph {et~al.}(2009)\citenamefont {Watanabe} \emph {et~al.}}]{Super-Kamiokande:2008mmn}%
  \BibitemOpen
  \bibfield  {author} {\bibinfo {author} {\bibfnamefont {H.}~\bibnamefont {Watanabe}} \emph {et~al.} (\bibinfo {collaboration} {Super-Kamiokande}),\ }\href {\doibase 10.1016/j.astropartphys.2009.03.002} {\bibfield  {journal} {\bibinfo  {journal} {Astropart. Phys.}\ }\textbf {\bibinfo {volume} {31}},\ \bibinfo {pages} {320} (\bibinfo {year} {2009})},\ \Eprint {http://arxiv.org/abs/0811.0735} {arXiv:0811.0735 [hep-ex]} \BibitemShut {NoStop}%
\bibitem [{\citenamefont {Syvolap}\ \emph {et~al.}(2022)\citenamefont {Syvolap}, \citenamefont {Ruchayskiy},\ and\ \citenamefont {Boyarsky}}]{Syvolap:2019dat}%
  \BibitemOpen
  \bibfield  {author} {\bibinfo {author} {\bibfnamefont {V.}~\bibnamefont {Syvolap}}, \bibinfo {author} {\bibfnamefont {O.}~\bibnamefont {Ruchayskiy}}, \ and\ \bibinfo {author} {\bibfnamefont {A.}~\bibnamefont {Boyarsky}},\ }\href {\doibase 10.1103/PhysRevD.106.015017} {\bibfield  {journal} {\bibinfo  {journal} {Phys. Rev. D}\ }\textbf {\bibinfo {volume} {106}},\ \bibinfo {pages} {015017} (\bibinfo {year} {2022})},\ \Eprint {http://arxiv.org/abs/1909.06320} {arXiv:1909.06320 [hep-ph]} \BibitemShut {NoStop}%
\bibitem [{\citenamefont {Schwartz}(2014)}]{Schwartz:2014sze}%
  \BibitemOpen
  \bibfield  {author} {\bibinfo {author} {\bibfnamefont {M.~D.}\ \bibnamefont {Schwartz}},\ }\href@noop {} {\emph {\bibinfo {title} {{Quantum Field Theory and the Standard Model}}}}\ (\bibinfo  {publisher} {Cambridge University Press},\ \bibinfo {year} {2014})\BibitemShut {NoStop}%
\end{thebibliography}%
\bibliographystyle{apsrev4-1}
\end{document}